\documentclass{article}
\usepackage{arxiv}

\usepackage[utf8]{inputenc} 
\usepackage[T1]{fontenc}    
\usepackage{hyperref}       
\usepackage{url}            
\usepackage{booktabs}       
\usepackage{amsfonts}       
\usepackage{nicefrac}       
\usepackage{microtype}      
\usepackage{lipsum}		
\usepackage{graphicx}
\usepackage{natbib}
\usepackage{doi}
\usepackage{cite}
\usepackage{amsmath,amssymb,amsfonts}
\usepackage{algorithmic}
\usepackage{graphicx}
\usepackage{float}
\usepackage{tabularx}
\usepackage{enumerate}
\usepackage{algorithm,algorithmic}
\usepackage{hyperref}
\usepackage{booktabs}
\usepackage{textcomp}
\usepackage{makecell}
\usepackage{multirow}
\allowdisplaybreaks

\title{Tuberculosis Screening from Cough Audio: \\Baseline Models, Clinical Variables, and Uncertainty Quantification}

\author{ \href{https://orcid.org/0000-0001-9701-1950}{\includegraphics[scale=0.06]{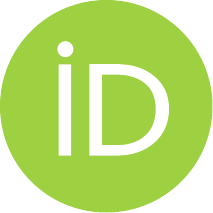}\hspace{1mm}George P. Kafentzis} \\
	Department of Computer Science\\
	University of Crete\\
	Greece \\
	\texttt{kafentz@csd.uoc.gr} \\
	\And
	{Efstratios Selisios} \\
	Department of Computer Science\\
	University of Crete\\
	Greece \\
	\texttt{selisios@csd.uoc.gr} \\
}

\renewcommand{\shorttitle}{\textit{arXiv} Template}

\hypersetup{
pdftitle={A template for the arxiv style},
pdfsubject={},
pdfauthor={Kafentzis, Selisios},
pdfkeywords={Tuberculosis, Machine Learning, Cough Audio, Cross-Validation, Uncertainty Quantification, Feature Extraction},
}

\begin{document}
\maketitle

\begin{abstract}
In this paper, we propose a standardized framework for automatic tuberculosis (TB) detection from cough audio and routinely collected clinical data using machine learning. While TB screening from audio has attracted growing interest, progress is difficult to measure because existing studies vary substantially in datasets, cohort definitions, feature representations, model families, validation protocols, and reported metrics. Consequently, reported gains are often not directly comparable, and it remains unclear whether improvements stem from modeling advances or from differences in data and evaluation. We address this gap by establishing a strong, well-documented baseline for TB prediction using cough recordings and accompanying clinical metadata from a recently compiled dataset from several countries. Our pipeline is reproducible end-to-end, covering feature extraction, multimodal fusion, cougher-independent evaluation, and uncertainty quantification, and it reports a consistent suite of clinically relevant metrics to enable fair comparison. We further quantify performance for cough audio-only and fused (audio + clinical metadata) models, and release the full experimental protocol to facilitate benchmarking. This baseline is intended to serve as a common reference point and to reduce methodological variance that currently holds back progress in the field. 
\end{abstract}


\section{Introduction}\label{intro}
Tuberculosis (TB) is a contagious disease caused by the bacterium Mycobacterium tuberculosis~\citet{CDCTB}. Although the lungs are its primary site of infection, TB can also affect other regions of the body, including the bones, lymph nodes, and brain. The disease spreads through airborne droplets when an infected individual coughs, sneezes, or speaks. Common symptoms of TB include persistent coughing, chest pain, fatigue, weight loss, and fever. Treatment typically involves a combination of antibiotics, but without proper intervention, TB can be fatal. Individuals with compromised immune systems, such as those living with HIV/AIDS, are particularly susceptible to developing TB. Globally, it is estimated that a quarter of the population is infected with TB, with over $10$ million new cases reported annually, although most of them remain latent~\citet{WHO-TB}.

The global eradication of TB is an urgent public health priority for several compelling reasons~\citet{Matteelli180035}. First, TB remains one of the leading causes of mortality worldwide, with an estimated $1.5$ million deaths recorded in $2020$. Second, the economic impact of TB is significant, as it reduces workforce productivity and increases healthcare expenditures. Third, TB disproportionately affects vulnerable groups, such as those living in poverty, individuals with HIV/AIDS, and Indigenous communities. Finally, the emergence of drug-resistant TB strains poses a severe challenge to global health, as these forms of the disease are far more difficult to treat. Consequently, eliminating TB is essential for saving lives, mitigating poverty, protecting at-risk populations, and preventing the spread of drug-resistant variants.

The diagnosis of TB typically involves multiple tests, including chest X-rays, tuberculin skin tests, and sputum analysis to detect the presence of the TB bacterium~\citet{Knechel09}. In certain cases, additional procedures, such as blood tests or biopsies, may also be required. Although a persistent cough is a hallmark symptom of TB, it is shared by various other conditions, complicating diagnosis based solely on auditory cues. However, vocal audio has proven useful in disease classification studies, particularly during the recent COVID-19 pandemic~\citet{erdougan2021covid, deshpande2020overview, hassan2020covid, coppock2021end}. Parameters of speech, such as phonation and vowel articulation, have been successfully employed in machine learning approaches to detect Parkinson's disease~\citet{braga2019automatic}. Additionally, respiratory diseases, including asthma bronchiale (AB), have been diagnosed through the analysis of cough sounds~\citet{hee2019development}. Cough sounds have also been utilized in the screening and diagnosis of pulmonary conditions such as AB, chronic obstructive pulmonary disease (COPD), and TB~\citet{8239338}.

However, audio data collection for such tasks is not trivial. First, obtaining high-quality audio recordings requires a controlled environment to minimize background noise and ensure the clarity of relevant features, which may not always be feasible in clinical or field settings. Second, variability in patient compliance and differences in recording equipment can lead to inconsistencies in the data. Third, ethical and privacy concerns regarding the collection, storage, and use of patient audio data necessitate rigorous consent processes and secure data handling protocols. Additionally, the dataset must be representative of diverse demographics, including variations in age, gender, and language to ensure generalization of the models. Finally, the presence of comorbid conditions and the overlapping acoustic features of different diseases can further complicate the labeling and analysis of the collected data, introducing potential biases and reducing model accuracy. Addressing these obstacles requires interdisciplinary collaboration, robust data collection methodologies, and ethical frameworks to enable the effective use of audio data in clinical applications. In our work, we obtain cough audio signals from a recently compiled dataset of cough sounds from individuals with and without TB~\citet{Huddart2024SolicitedCoughTB}. This dataset comprises $733,756$ cough recordings from $2143$ patients in seven countries, accompanied by detailed demographic, clinical, and microbiological diagnostic annotations. This dataset was originally created to support and assess the COugh Diagnostic Algorithm (CODA) for the TB DREAM Challenge~\citet{CODA}, which encouraged participants to develop algorithms for predicting TB diagnoses. The training data are now publicly accessible for broader use, organized in two sets: solicited (approximately $10,000$ coughs) and longitudinal coughs. To the best of our knowledge, this is one of the most diverse, rich in samples, high-quality, and publicly-available cough audio datasets that exist, thus making it suitable for our audio-based TB detection task. It should be noted that in this dataset, each patient may have more than one cough recording. 

Furthermore, on top of dataset variability, prior work also differs markedly in feature extraction and performance reporting, which further complicates interpretation. On the feature side, studies span engineered descriptors (e.g., Mel-Frequency Cepstral Coefficients (MFCCs)~\citet{Botha_2018, Pahar2021, Pahar2022} and other cepstral features~\citet{Yellapu2023}, spectral/energy statistics~\citet{Xu2023}, advanced spectrotemporal representations~\citet{Sharma2024}) 
 as well as learned representations such as pretrained audio embeddings~\citet{Pahar2022b} or end-to-end spectrogram-based models~\citet{Yadav2024}. The choice of machine-learning model is another major axis of variability across TB screening studies, spanning a wide spectrum of complexity. Classical tabular learners, such as Logistic Regression~\citet{Botha_2018}, support vector machines~\citet{Jember2023}, k-nearest neighbors~\citet{Mahmood2024}, random forests~\citet{Mahmood2024}, boosting methods and shallow neural networks~\citet{kafentzis2023predicting}, are commonly paired with engineered acoustic descriptors and clinical variables, often offering strong baselines with limited data and improved interpretability. In parallel, deep-learning approaches have been explored to learn representations directly from time--frequency inputs, including 1D and 2D convolutional neural networks (CNNs) on log-mel spectrograms~\citet{Yadav2024}, temporal models that capture cough dynamics~\citet{frost22_interspeech, Xu2023, Xu2024}, and transfer learning using pretrained audio encoders~\citet{Pahar2022, Bao2023, Yadav2024}. All of these choices are often coupled with different preprocessing decisions---segmentation strategy, denoising, amplitude normalization, silence trimming, and augmentation---that can substantially alter the information available to the classifier. On the reporting side, papers adopt an inconsistent mix of metrics (accuracy, F1-score, sensitivity/specificity, {area under the receiver operating characteristic curve} (ROC-AUC), and operating point measures such as sensitivity at a fixed specificity), sometimes without stating the decision threshold selection procedure. Because screening is inherently an operating point problem, this heterogeneity makes it difficult to translate numbers into clinically meaningful comparisons. In our work, we propose a collection of well-known engineered features, comprising Mel-Frequency Cepstral Coefficients, Chroma features, and simple spectral features, obtained on a frame-by-frame basis. Features are then statistically summarized over time using distributional summary statistics (moments and percentiles). The resulting vector is fed (with or without its corresponding clinical data) to a simple linear model, the well-known Logistic Regression~\citet{esl}, and a strong non-linear learner, CatBoost~\citet{prokhorenkova2018catboost}, a machine learning (ML) algorithm for gradient boosting on decision trees, i.e., an ensemble method that combines many simple if--then decision rules, added sequentially to improve accuracy, into a single final prediction. Each model is evaluated on a series of well-known metrics, namely ROC-AUC, area under the precision--recall characteristics curve (PR-AUC), Sensitivity, Specificity, UAR, PPV, and NPV. Details are presented in Section~\ref{methodology}.

Most importantly, studies also vary in data splitting protocols, and differences here can dominate reported performance. A common pitfall is evaluating on random splits at the recording level, which can inadvertently place coughs from the same participant in both training and test sets. Since cough acoustics include strong cougher- and device-specific signatures (e.g., habitual cough patterns, vocal tract traits, channel/microphone characteristics), this leakage can inflate performance and overstate generalization. For TB screening, the realistic target is prediction on previously unseen subjects, so evaluation should enforce cougher/subject independence via grouped cross-validation by cougher, leave-one-cougher-out validation, or an explicit cougher-disjoint held-out test set. When multiple recordings per subject or repeated sessions are present, protocols should also clarify how calibration/threshold selection is performed. Standardizing cougher-disjoint evaluation is therefore essential for fair benchmarking and for identifying improvements that truly generalize beyond the individuals seen during training. In our work, a cougher-independent, stratified, grouped, $10$-fold outer by $5$-fold inner, cross-validation strategy \mbox{is followed.}

Even when a model achieves strong average performance, its per-sample confidence can vary dramatically, especially in cough-based screening where recording conditions, background noise, co-morbidities, and symptom heterogeneity introduce substantial uncertainty. In a high-stakes triage setting, a single hard label is often insufficient: clinicians benefit from knowing \textit{how reliable 
} a prediction is, whether a case is an ``easy'' one or borderline, and when the system should defer and recommend confirmatory testing. Importantly, common confidence surrogates---such as the raw score output (often confused with an actual probability) of a classifier---are frequently \textit{miscalibrated}, meaning that high predicted scores do not necessarily correspond to high empirical correctness. This mismatch can lead to overconfidence errors. As a result, uncertainty-aware screening systems increasingly aim to provide \textit{interpretable confidence information}, support risk-stratified decision thresholds, and identify low-confidence cases for retesting or referral. Accordingly, we apply probability calibration (via isotonic regression~\citet{isotonic}, a monotone nonparametric mapping of scores to calibrated probabilities) to improve the interpretability of the model’s outputs as absolute risk probabilities. In addition, to quantify predictive confidence in a principled and \textit{model-agnostic} way, we adopt conformal prediction (CP)~\citet{JMLR:v9:shafer08a, molnar2023conformalpython}, which wraps around any trained classifier and produces statistically grounded measures of uncertainty. Conformal methods use a held-out calibration set to transform model scores into \textit{prediction sets} (or, in binary tasks, confidence/credibility measures) with a user-chosen coverage guarantee: specifically, under standard exchangeability assumptions, conformal prediction ensures \emph{finite-sample marginal coverage}, i.e., the prediction set contains the true label with probability at least $1-\alpha$. In practice, this yields actionable behavior for TB screening: confident cases may receive a single-label prediction, while ambiguous cases return a two-label set (or trigger abstention), explicitly signaling uncertainty and supporting safer decisions. Because conformal prediction does not assume a particular model family, it provides a consistent framework for reporting and comparing uncertainty across methods alongside conventional performance metrics. In our experiments, we have repeated recordings per cougher, and thus, we implement CP at the cougher level (via probability aggregation) to better align the exchangeability unit with evaluation.

Our main contributions are the following:
\begin{itemize}
    \item Setting up a standardized pipeline for training two representative (but practically any) machine learning models on the solicited part of the CODA TB dataset, comprising $9772$ cough audio samples from $1105$ individuals (both TB-positive and TB-negative). 
    \item Applying a cougher-independent cross-validation strategy that prevents information leakage between sets: we apply a $10$-by-$5$ nested, cougher-disjoint, cross-validation scheme, along with a disjoint conformal calibration subset used for operating point selection and conformal calibration (with probability calibration applied post-hoc).
    \item Examining acoustic features and fusing them with clinical features, demonstrating their capacity on predicting TB.
    \item Reporting a variety of evaluation metrics on both waveform and, most importantly, cougher/participant level.
    \item Quantifying model uncertainty using conformal prediction.
\end{itemize}

 We emphasize that the proposed baseline is intentionally model-agnostic: although two reference models are introduced, the core of our contribution lies in the specification of a leakage-free evaluation protocol rather than a particular feature representation or classifier. Consequently, it applies equally to approaches based on engineered features and to deep learning methods that learn representations end-to-end from waveforms or time--frequency inputs. To ensure fair comparison, all representation learning and hyperparameter tuning are performed strictly within the training portion of each cross-validation fold, with evaluation and uncertainty quantification carried out on cougher-independent held-out subjects.

The rest of this paper is organized as follows. Section~\ref{literature} discusses related work on TB vs. non-TB detection tasks from audio signals. Section~\ref{methodology} presents our approach, highlights key points, and visualizes our modeling pipeline. In addition, Section~\ref{experiments} summarizes our results and stresses the importance of each part of the pipeline, and Section~\ref{discussion} provides a discussion of the results. Finally, Section~\ref{conclusions} concludes the paper.

\section{Related Work} \label{literature}
Attempts to detect tuberculosis (TB) in an automatic way are not new~\citet{Hripcsak:97}, but the use of cough sounds for this purpose is relatively recent, fueled primarily by the COVID-19 pandemic~\citet{GeneralPaper:21, ScreenDiag:22}. Prior work explores a wide range of feature representations and model families, which is expected in an emerging task. However, reported performance is often hard to compare across studies because of two dominant sources of heterogeneity: \emph{(i) dataset variability} and \emph{(ii) evaluation protocol mismatch}.

\subsection{Representative Prior Approaches}
Early and mid-scale studies typically rely on private cohorts with tens of participants and varying numbers of cough segments. For example, in~\citet{Botha_2018}, the authors compiled a dataset of $38$ patients (TB/non-TB) and reported $95\%$ sensitivity at $72\%$ specificity using spectral and clinical characteristics. Pahar et al.~\citet{Pahar2021} used $51$ participants and trained conventional machine learning models on spectral/temporal features under nested cross-validation, achieving an ROC-AUC of $0.94$. Deep learning models have also been explored on mel-spectrograms and MFCCs, e.g.,~\citet{frost22_interspeech, Pahar2022, Pahar2022b, Jember2023}, reporting varying sensitivity/specificity and ROC-AUC values (Table~\ref{tab:cough_analysis}). In~\citet{Pahar2022}, the best-performing model was a pre-trained ResNet50 (a $50$-layer residual convolutional network~\citet{7780459}, initialized with weights learned on ImageNet and adapted via transfer learning), attaining an ROC-AUC of $0.92$. More recent works consider larger and/or more heterogeneous datasets, including other respiratory diseases and employ more complex neural architectures and fusion schemes~\citet{Yuan2023, Bao2023, Yellapu2023, Xu2023, Rajasekar2024, Mahmood2024, Xu2024, Sharma2024}. At the high-data end of the spectrum, ref.~\citet{Yadav2024}
 uses the extended CODA dataset (approximately $500{,}000$ solicited and longitudinal cough sounds) and reports an ROC-AUC of $0.91$ with a simple 1D convolutional neural network. Table~\ref{tab:cough_analysis} summarizes datasets, representations, and reported metrics.
 
 \subsection{Dataset Variability}
Across studies, datasets differ substantially in cohort size and composition, recording conditions, cough type (e.g., forced vs. passive), and label definitions and inclusion criteria. Many datasets are collected locally and may therefore reflect site- and population-specific factors. As a result, reported performance may not transfer to broader cohorts without cross-population validation.

\subsection{Protocol Mismatch and Potential Leakage}
A second major source of non-comparability is evaluation strategy. Studies vary between single train--test splits, $k$-fold cross-validation, and nested cross-validation, and these choices can materially affect performance estimates, especially on small datasets. Moreover, many works do not clearly state whether splits are \emph{grouped/subject-wise} (cougher-independent), i.e., whether all cough samples from the same individual are confined to a single fold/split. If cough samples from the same individual appear in both training and validation/test sets, performance can be biased upward due to subject-specific leakage.

\subsection{Our Proposal}
To address these issues, we exploit the largest publicly available cough dataset for TB, namely the CODA TB dataset, and evaluate it using a $10$-by-$5$ cougher-independent nested cross-validation protocol. This ensures that all samples from the same individual appear in exactly one of train/validation/test within each nested split. Finally, we apply conformal prediction as a model-agnostic uncertainty quantification framework to characterize the uncertainty of model predictions. 

\begin{table}[htb!]
\centering
\caption{\it Summary of studies on cough sound analysis for TB detection. The last two entries in this table present two publications directly related to the CODA TB dataset, but only~\citet{Rajasekar2024} uses the exact same dataset as in our work.}
\scriptsize
\begin{tabularx}{.9\textwidth}{cccccc}
\hline
\textbf{Author} & \textbf{Models} & \textbf{Dataset} & \textbf{Features} & \textbf{Metrics} & \textbf{Validation} \\ \hline \hline
(\citet{Botha_2018}) & Logistic Regression & 746 coughs & \makecell[c]{5 spectral and\\ clinical characteristics} & \makecell[c]{Sens: 0.95,\\ Spec: 0.72}  & nested $5-$fold CV \\ \hline
(\citet{Pahar2021}) & \makecell[c]{Conventional ML \\ + a CNN} & 1,358 coughs & \makecell[c]{23 spectral and \\ temporal features} & \makecell[c]{Sens: 0.93,\\ Spec: 0.95}  & nested $5-$fold CV\\ \hline
(\citet{frost22_interspeech}) & Bi-LSTMs & 1,564 coughs & Mel-spectrograms & \makecell[c]{Sens: 0.89,\\ Spec: 0.75} & \makecell[c]{single train-test split \\ with $4-$fold CV} \\ \hline
(\citet{Pahar2022}) & \makecell[c]{CNN, LSTM \\ ResNet50 \\ + transfer learning} & $>6000$ coughs & MFCCs & \makecell[c]{Sens: 0.96,\\ Spec: 0.80} & nested $5-$fold CV \\ \hline
(\citet{Pahar2022b}) & \makecell[c]{Shallow and deep \\ learners} & 1,235 coughs & \makecell[c]{NLP-style \\ cough embeddings} & AUC: 0.81 & nested $5-$fold CV \\ \hline
(\citet{Jember2023}) & ANN, SVM & 3,238 coughs & MFCCs & F1: 0.87 & \makecell[c]{single train-test split \\ with $k$-fold CV} \\ \hline
(\citet{Yuan2023}) & \makecell[c]{CNN, dual-branch \\ model} & 1,323 coughs & \makecell[c]{MFCCs, \\ Mel-spectrograms} & \makecell[c]{Sens: 0.91, \\ AUC: 0.93, \\ F1: 0.91} & single $60-20-20$ split \\ \hline
(\citet{Bao2023}) & \makecell[c]{GoogleNet, \\ ResNet50, \\ Fusion model} & 1,000 coughs & MFCCs & \makecell[c]{Sens: 0.989, \\ Spec: 0.975}  & \makecell[c]{single $8:2$ split with \\$10$-fold CV} \\ \hline
(\citet{Yellapu2023}) & \makecell[c]{Combination of ANN \\ and CNN} & 3,102 coughs & \makecell[c]{170 spectral \\ and temporal \\ features} & \makecell[c]{Sens: 0.903, \\ Spec: 0.847} & \makecell[c]{single $8:2$ split with \\$10$-fold CV}\\ \hline
(\citet{Xu2023}) & \makecell[c]{Dynamic convolutions, \\ self-attention} & 1,323 coughs & Spectral features & \makecell[c]{Sens: 0.977, \\ Spec: 0.994} & single $60-20-20$ split \\ \hline
(\citet{Mahmood2024}) & Conventional ML models & 870 coughs & \makecell[c]{206 temporal and \\ spectral features} & \makecell[c]{Sens: 1.00, \\ F1: 0.97} & single $67-33$ split\\ \hline
(\citet{Xu2024}) & \makecell[c]{Bi-LSTM, Bi-GRU, \\ CNN} & 456 coughs & \makecell[c]{Spectral and \\ temporal features} & \makecell[c]{Sens: 0.98, \\ Spec: 0.95}  & single $80-20$ split \\ \hline
(\citet{Sharma2024}) & ResNet18 & 34,866 coughs & Scalograms & \makecell[c]{Sens: 0.70, \\ Spec: 0.71, \\ AUC: 0.79}  & $5-$fold CV  \\ \hline
(\citet{Rajasekar2024}) & Capsule networks & 9,772 coughs & \makecell[c]{Spectral and \\ HOG features} & \makecell[c]{Sens: 0.98, \\ Spec: 0.96, \\ F1: 0.97} & single $80-20$ split\\ \hline
(\citet{Yadav2024}) & \makecell[c]{1D \& 2D-CNNs, \\ ResNet50, VGG16} & \makecell[c]{502,252 \\ coughs} & \makecell[c]{Mel-spectrograms, \\ MFCCs} & \makecell[c]{Sens: 0.84, \\ F1: 0.87, \\ AUC: 0.91} & $5-$ fold CV\\ \hline
\end{tabularx}
\label{tab:cough_analysis}
\end{table}

From this review, it is evident that there is a large number of features, classifiers, and datasets that have been used to detect TB from audio. At the same time, it is hard to compare different approaches due to their differences in (a) datasets, and (b) validation strategy. Private datasets or datasets obtained from local populations potentially introduce systematic biases influenced by regional environmental, genetic, and socio-cultural factors. Thus, caution must be exercised when generalizing these findings to more diverse or international cohorts, highlighting the need for broader, cross-population validation studies. Validation strategies that include single train-test splits may yield an unreliable estimate of the performance of a model, as the evaluation is highly dependent on the specific characteristics of the chosen split, which may not represent the true variability of the data. This approach increases the risk of overfitting and limits the generalizability of the model, highlighting the importance of employing more robust validation techniques such as cross-validation. Moreover, in the majority of these works, there is no clear mention about whether the split or the cross-validation is grouped (or cougher-independent), that is, cough samples from a specific individual are all kept either in the training or in the validation/test set but not in both. In the latter case, the results are biased since the model has been trained on cough samples from a cougher that also has samples in the validation/test set, leading to artificially inflated performance.

To tackle these problems, in our work we exploit the largest publicly available dataset of cough sounds, namely the CODA TB dataset, while we validate our approach using a $10$ by $5$, cougher independent, nested cross-validation, ensuring that all samples from the same cougher are either in the train, the validation set, the test set, or any other mentioned set, but not in any two, three, or in all of them. Moreover, we apply an uncertainty quantification framework using a model-agnostic methodology named conformal prediction to examine the uncertainty of model predictions. 

\section{Methodology}
\label{methodology}
In this Section, we thoroughly present our methodology, roughly divided into five modules; first, we briefly present the CODA dataset followed by the splitting strategy of the pipeline. Second, the feature extraction process is discussed, and third, we briefly state a short description of the models used in this work. Finally, we provide the calibration and thresholding scheme, as well as the evaluation metrics and the uncertainty quantification framework.

\subsection{CODA Dataset}
In this paragraph, we briefly present the dataset used in our work - for a more detailed description, the interested reader can see~(\citet{CODAweb, Huddart2024SolicitedCoughTB}). The dataset was collected at outpatient health centers in seven countries (India, the Philippines, South Africa, Uganda, Vietnam, Tanzania, and Madagascar). The study screened adults ($\geq 18$ years) who attended these clinics for any reason. Individuals reporting a new cough or a worsening cough lasting at least two weeks were enrolled.

At the enrollment visit, participants completed a questionnaire capturing standard demographic and clinical variables, and a sputum specimen was obtained for tuberculosis (TB) testing. Participants were also asked to produce coughs that were recorded. The dataset contains a mixture of solicited and spontaneous coughs. All audio was recorded at a sampling rate of $44,100$~Hz, and was downsampled as required for individual experiments (details provided in subsequent sections). Each cough had a duration of $0.5$ seconds (very few recordings needed zero-padding to that length). In total, $9,772$ cough sounds were included. Descriptive statistics are summarized in Table~\ref{tab:stats} and Table~\ref{tab:metadata} lists demographic and clinical variables.

\begin{table}[htb!]
\centering
\caption{\it Summary statistics of the dataset. TB$+$ and TB$-$ denote participants with and without tuberculosis, respectively.}
{\small
\begin{tabularx}{0.6\textwidth}{llll}
\hline
\textbf{Metric} & \textbf{TB+} & \textbf{TB-} & \textbf{Total} \\ \hline\hline
Participants & 295 & 810 & 1105 \\ \hline
Total coughs & 2930 & 6842 & 9772 \\ \hline
\makecell[l]{Average number of\\ coughs/participant ($\pm$ std)} &
\makecell[l]{10.06\\ ($\pm$ 6.48)} &
\makecell[l]{8.65\\ ($\pm$ 5.15)} &
\makecell[l]{9.03\\ ($\pm$ 5.7)} \\ \hline
Minimum number of coughs/participant & 3 & 3 & -- \\ \hline
Maximum number of coughs/participant & 50 & 37 & -- \\ \hline
Total duration of coughs (minutes) & 24.41 & 57.01 & 81.43 \\ \hline
\end{tabularx}}
\label{tab:stats}
\end{table}


It should be noted that this work is not the first to utilize this dataset. Similarly to our work, ~(\citet{Rajasekar2024}) used part of the CODA dataset, in the form of spectral and HOG (histogram-of-gradients) features, to train convolutional and capsule networks, resulting in an F1 score of $0.97$, sensitivity of $0.98$, and specificity of $0.96$, in a single $80-20$ stratified split, but without mentioning whether the splits were cougher-disjoint. In addition,~(\citet{Yadav2024}) transformed the entire CODA dataset (both solicited and longitudinal cough sounds) into MFCCs and mel-spectrograms to train a 1D and a 2D convolutional neural networks, along with two pretrained models (ResNet50, VGG16), in a 5-fold cross-validation, yielding an average ROC AUC of $0.91$. However, no mention of cougher-disjoint splits is present in their work. Finally, the first author of this work along with other researchers had presented some results of the CODA Dream Challenge~(\citet{CODAweb}) which was based only on ROC AUC as an evaluation metric, using a similar $10$-by-$5$ stratified, grouped-by-cougher, CV scheme~(\citet{kafentzis2023predicting}).
\begin{table}[htb!]
\centering
\caption{\it Demographic and clinical variables used as features in the fused features experiment.}
{\small
\begin{tabularx}{\textwidth}{lll}
\hline
\textbf{Clinical or demographic datum} & \textbf{Description} & \textbf{Unit of measurement} \\ \hline \hline
Age & \makecell[l]{Computed as (collection date -- date of birth) when available;\\ otherwise, age reported at the time of collection.} & Years \\ \hline
Sex & Sex assigned at birth as reported by the participant & \makecell[l]{Binary\\ (male/female)} \\ \hline
Height & Participant height & Centimeters \\ \hline
Weight & Participant weight & Kilograms \\ \hline
Reported duration of coughing & Self-reported duration of the current cough & Days \\ \hline
Prior TB & Self-reported history of having TB or being told they had TB & \makecell[l]{Binary\\ (yes/no)} \\ \hline
Prior TB (Pulmonary) & Self-reported history of pulmonary TB & \makecell[l]{Binary\\ (yes/no)} \\ \hline
Prior TB (Extrapulmonary) & Self-reported history of extrapulmonary TB & \makecell[l]{Binary\\ (yes/no)} \\ \hline
Prior TB (Unknown) & \makecell[l]{Self-reported history of TB where pulmonary/extrapulmonary\\ status is not specified.} & \makecell[l]{Binary\\ (yes/no)} \\ \hline
Hemoptysis & Self-reported history of coughing up blood & \makecell[l]{Binary\\ (yes/no)} \\ \hline
Heart Rate & Baseline heart rate measurement & Beats per minute \\ \hline
Temperature & Baseline body temperature measurement & Celsius \\ \hline
Smoke last week & \makecell[l]{Self-reported use of combustible tobacco and/or vaping products\\ within the previous 7 days.} & \makecell[l]{Binary\\ (yes/no)} \\ \hline
Fever & \makecell[l]{Self-reported fever symptoms within the previous 30 days.} & \makecell[l]{Binary\\ (yes/no)} \\ \hline
Night sweats & \makecell[l]{Self-reported night sweats within the previous 30 days.} & \makecell[l]{Binary\\ (yes/no)} \\ \hline
Weight loss & \makecell[l]{Self-reported weight loss within the previous 30 days.} & \makecell[l]{Binary\\ (yes/no)} \\ \hline
\end{tabularx}}
\label{tab:metadata}
\end{table}

\subsection{Validation Strategy}
In our implementation, we follow a \textit{nested, cougher-disjoint cross-validation strategy}. 
At the top level, we use an outer stratified, cougher-grouped $10$-fold split, where the group corresponds to the subject (cougher) and the folds are shuffled with a fixed seed. Stratification is performed with respect to the TB label (TB+, TB$-$) at the cougher level. In each outer iteration, \textit{all samples from the held-out coughers form the test set}, while all remaining coughers constitute the outer training pool, ensuring cougher-independence at \mbox{evaluation time.}
 
Within each outer training pool, we further create a \textit{cougher-disjoint conformal calibration subset} by splitting coughers (not samples). This subset is reserved exclusively for post-training calibration steps: it is used to compute (i) the \textit{Youden threshold $\tau$ (a standardized baseline operating point)}~\citet{Youden1950Index}, and (ii) the \textit{conformal quantiles} $\hat{q}(\alpha)$. The remaining coughers form the proper-training subset used for (a) model hyperparameter selection and (b)~learning the predictive model and probability-calibration mapping.

Hyperparameter selection is performed only on the proper-training coughers via an \textit{inner 5-fold, stratified, cougher-grouped split}. Since the dataset is class-imbalanced, we select hyperparameters by maximizing the mean UAR (unweighted average recall or balanced accuracy) across inner folds. In each inner fold, UAR is computed after choosing a fold-specific operating threshold via Youden's criterion, which aligns model selection with the screening objective of balanced sensitivity and specificity rather than threshold-free ranking alone. While selecting $\tau$ on each inner validation fold can modestly inflate inner-fold UAR, this affects only hyperparameter ranking; the final operating threshold $\tau$ used for evaluation is selected on a disjoint calibration subset within the outer training pool.

\begin{figure*}[htb!]
    \centering
    \includegraphics[width=1\linewidth]{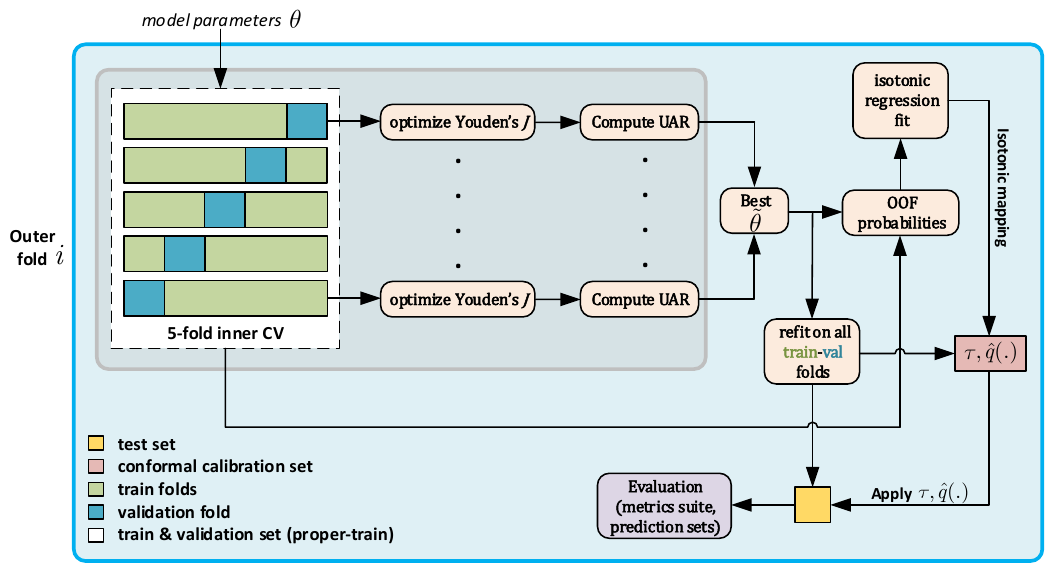}
    \caption{\it Cougher-disjoint nested CV pipeline for model selection, calibration, and conformal prediction based uncertainty quantification.}
    \label{fig:cv}
\end{figure*}

Successively, within each outer fold, probability calibration is learned using training data only. After selecting hyperparameters on the proper-training subset, we compute out-of-fold (OOF) probabilities on the proper-training subset using the selected hyperparameters and fit an isotonic regression mapping on these OOF predictions. We then refit the classifier on the full proper-training subset, obtain raw probabilities for the disjoint conformal calibration subset, and transform them using the isotonic mapping to yield calibrated probabilities. Decision thresholds $\tau$ (e.g., Youden's operating point) and conformal quantiles $\hat{q}(\alpha)$ are estimated exclusively from this conformal calibration subset and are subsequently applied to the calibrated probabilities of the held-out outer-test coughers for final evaluation. Figure~\ref{fig:cv} depicts the set splitting process for one fold.

\subsection{Feature Extraction}
\begin{figure}[htb!]
    \centering
    \includegraphics[width=.9\linewidth]{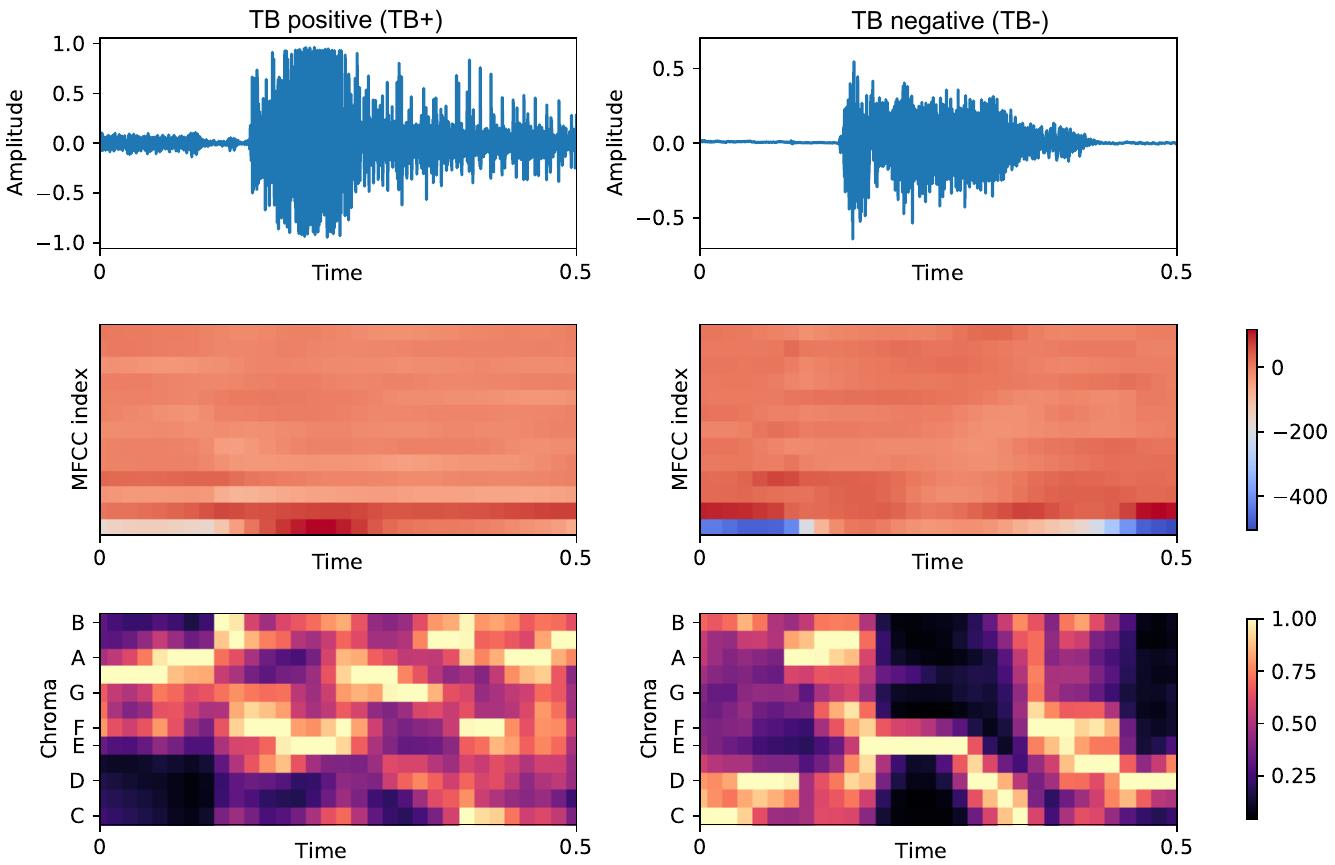}
    \caption{\it MFCC and Chroma features for two cough waveforms, TB+ and TB-.}
    \label{fig:feats}
\end{figure}

The literature indicates a wide range of compact feature representations for modeling sound signals, but most were originally designed for audio, music, or speech applications rather than cough-specific analysis. Cough is highly \textit{non-stationary}---its characteristics change rapidly over time---thus it is typically analyzed using short time segments (frames) that can be treated as approximately stationary. In this context, stationarity means temporal and spectral properties of a frame remain nearly constant within that interval. The frame duration must balance two needs: it should be long enough to estimate features reliably, yet short enough that the extracted features still reflect the content of that frame. For this reason, common frame lengths are on the order of 20--50 ms. Each frame is then tapered by multiplying it with a Hanning window, which reduces spectral artifacts and improves frequency-domain behavior. 
To avoid the dimensionality curse {(i.e., increased overfitting risk and sample requirements as feature dimension increases)~\citet{Altman2018CurseDimensionality}}, we would like to focus on features that are relatively low in number but sufficiently representative of the cough audio signal. There seems to be an agreement that spectral features are useful up to a certain level~\citet{peeters2004large}, and we list a selection of them below.
\begin{enumerate}
\item \textbf{Spectral Centroid:} The spectral centroid (SC) represents the spectrum’s ``center of mass'' within a frame. Interpreting the magnitude spectrum as a weighted distribution over frequency bins, the centroid is the weighted average of bin indices (or frequencies), with weights given by the spectral magnitudes. Let $X_n[k]$ denote the $N$-point FFT magnitude spectrum of the $n$th frame and
$f[k]$ be the center frequency (in Hz) of bin $k$. The centroid can be written as
\begin{equation}
C_n = \frac{\sum_{k=0}^{N-1} f[k] X_n[k]}{\sum_{k=0}^{N-1} X_n[k]}
\end{equation}
\item \textbf{Spectral Roll-off:} The $85\%$ spectral roll-off is the frequency (or bin) below which $85\%$ of the frame’s spectral energy is accumulated; equivalently, it is the $0.85$ quantile of the cumulative energy distribution.
\item \textbf{Spectral Bandwidth}: spectral bandwidth is the $p$th-order moment of the spectrum around the spectral centroid. Let $X_n[k]$ be the magnitude spectrum of the $n$th frame, and let
$f[k]$ be the center frequency (in Hz) of bin $k$. If $C_n$ denotes the spectral centroid at frame $n$, then the spectral bandwidth is
\begin{equation}
B_n = \left(\sum_{k=0}^{N-1} X_n[k]\left|f[k]-C_n\right|^{p}\right)^{\frac{1}{p}} .
\end{equation}
\item \textbf{Spectral Flatness}: spectral flatness quantifies how noise-like a spectrum is. It is computed as the ratio of the geometric mean to the arithmetic mean of a thresholded (and exponentiated)
spectrum. Let $X_n[k]$ be the magnitude spectrum of the $n$th frame. Then, the per-frame geometric and arithmetic means are
\begin{align}
G_n &= \exp\!\left(\frac{1}{N}\sum_{k=0}^{N-1}\log X_n[k]\right) \\
A_n &= \frac{1}{N}\sum_{k=0}^{N-1} X_n[k]
\end{align}
and the spectral flatness of the $n$th frame is
\begin{equation}
F_n = \frac{G_n}{A_n}
\end{equation}
\item \textbf{Mel-frequency cepstral coefficients---MFCCs:} {MFCCs are features computed by established signal-processing transforms, as opposed to learned representations obtained via end-to-end training.} They describe the short-term spectral envelope using a perceptually motivated frequency axis (the mel scale)~\citet{RabinerSchafer2011}. MFCC extraction is typically performed by (i) computing the power spectrum of each frame, (ii) integrating it through a bank of overlapping filters spaced on the mel scale to obtain mel-band energies, (iii) taking the logarithm to form log-mel filterbank energies, and (iv) applying a discrete cosine transform (DCT) to these log energies. The resulting DCT coefficients constitute the MFCCs.
\item \textbf{Chroma features:} Chroma features (chromagrams) summarize short-time spectral energy by folding frequencies into pitch classes (typically 12 bins), i.e., they emphasize where energy lies on a pitch-class axis {while being largely octave-invariant}~\citet{Muller2015FMP}. For TB cough detection, chroma is not a “natural” cough feature (cough is often broadband and aperiodic), but it can still be complementary: MFCCs primarily encode the spectral envelope (timbre/shape) while chroma can act as a compact descriptor of quasi-harmonic/voiced or resonant structure when present (e.g., laryngeal excitation components, repeated resonant bands) and add invariances that MFCCs do not explicitly target. This is not just hypothetical---at least one TB cough study~\citet{Xu2024} explicitly extracts Chroma features and uses it in a feature-fusion pipeline for pulmonary TB detection from cough sounds, suggesting it can contribute signal beyond standard spectral features in some settings.
\end{enumerate}
An example of MFCC and Chroma feature extraction for two coughs, a TB-positive and a TB-negative, is depicted in Figure~\ref{fig:feats}.
Table~\ref{tab:feat_params} lists the parameters for feature extraction. In total, each frame is encoded by $4$ spectral features, $13$ MFCCs, and $12$ Chroma features, resulting in {$M=29$} features per frame, over {$L=32$} overlapping frames.
\begin{table*}[t]
\centering
\caption{\it Feature extraction configuration for cough audio.}
\label{tab:feat_params}
\small
\begin{tabular}{l l}
\hline
\textbf{Component} & \textbf{Parameters} \\
\hline \hline
Sampling Rate & $16000$ Hz\\ \hline
MFCCs & \makecell[l]{$N_{\mathrm{FFT}}=2048$ bins \\ \texttt{Window length}$=32$ ms \\ \texttt{frame rate}$=16$ ms\\$N_{\mathrm{MFCC}}=13$}\\ \hline
Chroma features & \makecell[l]{$12$ chroma bins \\ $N_{\mathrm{FFT}}=2048$ bins \\ \texttt{Window length}$=32$ ms \\ \texttt{frame rate}$=16$ ms}\\ \hline
\makecell[l]{Spectral centroid\\Spectral bandwidth\\Spectral roll-off ($85\%$)\\Spectral flatness} & \makecell[l]{$N_{\mathrm{FFT}}=2048$ bins \\ \texttt{Window length}$=32$ ms \\ \texttt{frame rate}$=16$ ms}\\
\hline
\end{tabular}
\end{table*}
A set of statistical functionals is applied to these features to convert the frame-level sequences into a fixed-length representation per recording. Let 
 $\mathbf{x}=\{x_n\}_{n=1}^L = [x_1,x_2,\ldots,x_L]^T$ denote an example feature trajectory over time, consisting of $L$ time instants (frames). We compute:
\begin{enumerate}
\item Mean: the sample average of a feature
\begin{equation}
\mu = \frac{1}{L}\sum_{n=1}^L x_n
\end{equation}
\item Standard deviation: a measure of dispersion around the mean
\begin{equation}
\sigma = \sqrt{\frac{1}{L-1}\sum_{n=1}^L (x_n-\mu)^2}
\end{equation}
\item Skewness: quantifies the asymmetry of the empirical distribution about its mean
\begin{equation}
\gamma_1 = \frac{\sqrt{L(L-1)}}{L-2}\frac{\frac{1}{L}\sum_{n=1}^L (x_n-\mu)^3}{\left(\frac{1}{L}\sum_{n=1}^L (x_n-\mu)^2\right)^{3/2}}
\end{equation}
\item Kurtosis: describes the heaviness of the tails of the empirical distribution
\begin{align}
\gamma_2 &= \frac{(L+1)L}{(L-1)^3(L-2)(L-3)} \cdot \frac{\sum_{n=1}^L (x_n-\mu)^4}{\sigma^4}
- 3\frac{(L-1)^2}{(L-2)(L-3)}
\end{align}
\item The 10th percentile ($P_{10}$): the value below which $10\%$ of the samples fall;
\item The 25th percentile ($P_{25}$): the first quartile ($Q_1$);
\item The 50th percentile ($P_{50}$): the median (second quartile, $Q_2$);
\item The 75th percentile ($P_{75}$): the third quartile ($Q_3$);
\item The 90th percentile ($P_{90}$): the value below which $90\%$ of the samples fall.
\end{enumerate}
Now, the $m${th} feature trajectory over time, with $m=1,\ldots, M=29$, is compactly represented by these nine summary statistics as
\begin{equation}
    \mathbf{F}_m = \left[\mu, \sigma, \gamma_1, \gamma_2, P_{10}, P_{25}, P_{50}, P_{75}, P_{90} \right]^T
\end{equation}
and the final feature vector for an entire recording is
\begin{equation}
    \mathbf{V} = \left[\mathbf{F}_1^T, \mathbf{F}_2^T, \ldots \mathbf{F}^T_m, \ldots, \mathbf{F}_M^T \right]^T
\end{equation}
where $^T$ denotes transposition. Feature vector $\mathbf{V}$ is $29\times 9 =261$-dimensional for each audio recording. The proposed feature set choice is motivated by (a) straightforward implementation (often available in open-source toolkits such as LibROSA~\citet{librosa}), (b) robustness, (c) relevance to the task, and (d) suitability for near real-time processing.

For clinical data, features were encoded as follows: continuous-valued measurements (e.g., age, height, weight, reported cough duration, heart rate, temperature) were retained as real-valued features, while binary variables (sex and yes/no symptom indicators) were encoded as $0/1$. No missing values were found in the dataset. Features were standardized using z-score normalization with scaling parameters fit on training data and applied to held-out splits.

\subsection{Machine Learning Models}
Although we could have tested an extensive set of ML models, ranging from simple linear models to powerful non-linear ensembles, only two models were selected: Logistic Regression (LR)~\citet{stoltzfus2011logistic}, a very simple, well-known linear model, used by many ML practitioners and present in many research works around TB detection from audio (see Table~\ref{tab:cough_analysis}), and CatBoost (CB)~\citet{prokhorenkova2018catboost}, a gradient-boosting algorithm that builds an ensemble of decision trees while using an ordered target-statistics scheme for categorical features to reduce target leakage and prediction shift, typically yielding strong performance with minimal preprocessing. LR has been selected for its wide adoption and simplicity while CB has been chosen for its great performance on tabular data~\citet{erickson2025tabarena, shmuel2024comprehensive}. 

In brief, LR models the conditional probability that a participant is tuberculosis-positive
(\(y=1\)) given a \(d\)-dimensional feature vector
\begin{equation}
\mathbf{V}=[v_1,v_2,\ldots,v_d]^T\in\mathbb{R}^d.
\end{equation}
To 
 include an intercept, we define the augmented vector \(\tilde{\mathbf{V}}=[1,\mathbf{V}^T]^T\in\mathbb{R}^{d+1}\)
and parameter vector \(\boldsymbol{\theta}=[\theta_0,\theta_1,\ldots,\theta_d]^T\in\mathbb{R}^{d+1}\).
LR assumes that the log-odds are linear in \mbox{the features,}
\begin{equation}
\log\frac{P_{TB}(\mathbf{V})}{1-P_{TB}(\mathbf{V})}=\boldsymbol{\theta}^T\tilde{\mathbf{V}},
\end{equation}
which is equivalent to the sigmoid parameterization
\begin{equation}
P_{TB}(\mathbf{V})=\sigma(\boldsymbol{\theta}^T\tilde{\mathbf{V}})
=\frac{1}{1+\exp\!\big(-\boldsymbol{\theta}^T\tilde{\mathbf{V}}\big)}.
\end{equation}
Given a training dataset $D = \{(\mathbf{V}_i,y_i)\}_{i=1}^n$ with \(y_i\in\{0,1\}\), parameters are typically estimated by maximizing penalized (l2) likelihood
\begin{equation}
\boldsymbol{\theta}=\arg\max_{\boldsymbol{\theta}} \ell(\boldsymbol{\theta})
\end{equation}
where
\begin{equation}
\qquad \ell(\boldsymbol{\theta})=\sum_{i=1}^n \Big[y_i\log p_i+(1-y_i)\log(1-p_i)\Big]-\frac{1}{2C}||\theta||_2^2
\end{equation}
with $C$ being the inverse regularization strength, and \(p_i=P_{TB}(\mathbf{V}_i)\). We do not regularize the intercept $b$, consistent with standard implementations.

On the other hand, {CatBoost is a gradient-boosted decision-tree model trained on the dataset
$D=\{(\mathbf{V}_i,y_i)\}_{i=1}^{n}$, where $\mathbf{V}_i\in\mathbb{R}^{d}$ and $y_i\in\{0,1\}$.
It builds an additive predictor $F_t:\mathbb{R}^{d}\rightarrow\mathbb{R}$ over
boosting iterations $t=1,\ldots,T$:
\begin{equation}
F_t(\mathbf{V})=F_{t-1}(\mathbf{V})+\eta h_t(\mathbf{V})
\label{eq:catboost_update}
\end{equation}
where $\eta>0$ is the learning rate, and $h_t\in\mathcal{H}$ is a regression tree. Here, $\mathcal{H}$ denotes the family of regression trees permitted by the CatBoost configuration (e.g., constraints on tree depth and leaf regularization; see Table~\ref{tab:model_hparams_cb_lr}).
Given a differentiable loss $\ell(y,s)$ and the current score $s=F_{t-1}(\mathbf{V})$,
CatBoost computes pseudo-residuals (negative loss gradients) at the \mbox{training points:}
\begin{equation}
r_{t,k}=-\left.\frac{\partial \ell(y_k,s)}{\partial s}\right|_{s=F_{t-1}(\mathbf{V}_k)}
\label{eq:catboost_residuals}
\end{equation}
and fits $h_t$ to these targets, typically via least squares:
\begin{equation}
h_t=\arg\min_{h\in\mathcal{H}}\frac{1}{n}\sum_{k=1}^{n}\left(r_{t,k}-h(\mathbf{V}_k)\right)^2
\label{eq:catboost_ls}
\end{equation}
CatBoost additionally handles categorical predictors using target-based statistics (i.e.,
smoothed target encoding with regularization, commonly computer in an ordered manner to reduce target leakage)~\citet{prokhorenkova2018catboost}}. Table~\ref{tab:model_hparams_cb_lr} lists each model's fixed settings and their hyperparameter search space.
\begin{table*}[t]
\centering
\caption{\it Model hyperparameters and search spaces used in our implementation.}
\label{tab:model_hparams_cb_lr}
\small
\begin{tabular}{l p{7.2cm} p{6cm}}
\hline
\textbf{Item} & \textbf{CatBoost} & \textbf{Logistic Regression} \\
\hline \hline
Fixed settings &
\texttt{loss\_function="Logloss"}, \texttt{eval\_metric="AUC"}, \texttt{random\_seed=42}&
\texttt{max\_iter=10000}, \texttt{random\_state=42}. \\
\hline
\makecell[l]{Tuned \\hyperparameters} &
\begin{tabular}[t]{@{}l@{}}
\texttt{depth} $\in \{4,6,8\}$\\
\texttt{iterations} $\in \{400,800,1200\}$\\
\texttt{learning\_rate} $\in \{0.03,0.10\}$\\
\texttt{l2\_leaf\_reg} $\in \{1.0,3.0,10.0\}$\\
\texttt{subsample} $\in \{0.7,0.9,1.0\}$\\
\texttt{rsm} $\in \{0.7,0.9,1.0\}$\\
\texttt{auto\_class\_weights} $\in \{\texttt{None},\texttt{"Balanced"}\}$
\end{tabular}
&
\begin{tabular}[t]{@{}l@{}}
\texttt{inverse\_regularization\_strength\ (C)} \\
$\in \{10^{-4},5\cdot10^{-4},10^{-3},10^{-2},5\cdot10^{-2},10^{-1}\}$\\
\texttt{penalty} $=\texttt{"l2"}$\\
\texttt{solver} $\in \{\texttt{"liblinear"},\texttt{"lbfgs"}\}$\\
\texttt{class\_weight} $\in \{\texttt{None},\texttt{"balanced"}\}$
\end{tabular}
\\
\hline
\end{tabular}
\end{table*}

\subsection{Calibration and Thresholding}
{In our work, we explicitly treat probability calibration as a distinct post-hoc step from model fitting. This step is particularly important when outputs are used for risk assessment, since decision rules and clinical communication often rely on absolute risk estimates (e.g., “refer if risk is greater than $p$, where $p \in [0,1]$”), which are only meaningful if the reported probabilities are well-calibrated. After hyperparameter selection, we calibrate the model’s raw scores so that the resulting probabilities are better aligned with empirical correctness. We use \textit{isotonic regression}~\citet{isotonic}, a flexible monotone mapping that can correct systematic over- or under-confidence without assuming a parametric form. To avoid optimistic bias and information leakage, the calibrator is fit only on out-of-sample predictions from the \emph{proper-training} data (e.g., out-of-fold (OOF) predictions obtained within the outer-training partition), ensuring that a sample is not used to both train the scoring model and calibrate its own prediction. Within each outer fold, the resulting calibration function is then applied to probabilities produced on the disjoint calibration subset and on the held-out outer test subjects. While our primary binary operating point is chosen by maximizing Youden’s $J$ (which depends primarily on score ordering and is therefore largely invariant to monotone score transformations), calibrated probabilities remain valuable for (a) meaningful interpretation and communication of individual-level risk, and (b) enabling alternative probability-based decision rules (e.g., fixed risk cutoffs or cost/utility-based thresholds) should deployment requirements evolve.}

{We also separate \textit{decision thresholding} from probability calibration and choose operating points using a held-out \emph{calibration (conformal) subset} that is disjoint from the final test subjects. Rather than fixing an arbitrary threshold (such as $0.5$, commonly used in the literature), we select one that aligns with screening objectives; in particular, we use \textit{Youden’s index} to identify the threshold $\tau$ that maximizes the sensitivity--specificity trade-off. Because screening decisions may be made either per recording or per cougher, we estimate thresholds at both the waveform level, $\tau_w$, and the cougher level, $\tau_s$, after aggregating multiple recordings (e.g., by averaging probabilities). This design yields a leakage-free estimate of decision thresholds, while still reporting threshold-independent metrics (ROC-AUC, PR-AUC) for model comparison, and it makes explicit how operating choices influence clinically relevant measures such as sensitivity and specificity.}

To evaluate the reliability of probabilistic outputs, we report calibration diagnostics before and after isotonic regression. Within each outer fold, isotonic regression is fit on OOF probabilities obtained from the inner cross-validation performed on the proper-training partition, yielding a monotone mapping from raw to calibrated probabilities without using outer-test labels. Calibration is assessed on the outer test partition using the Brier score (BS)
\begin{equation}
    BS = \frac{1}{N}\sum_{i=1}^N (\hat{p}_i-y_i)^2
    \label{eq:brier}
\end{equation}
where $\hat{p}_i$ is the predicted probability for the positive class of sample $i$, and $y_i$ is the label of that sample and the expected calibration error (ECE), computed by binning predicted probabilities and measuring the absolute deviation between empirical accuracy and mean confidence per bin. {Specifically, we partition $[0,1]$ into $M=10$ equal-frequency bins. In each bin, $B_m$, $m=1,\cdots,10$, we compute the empirical event rate (empirical accuracy)
\begin{equation}
    \hat{\pi}_m = \frac{1}{|B_m|}\sum_{i \in B_m}y_i
\end{equation}}
{and the mean predicted probability (mean confidence)
\begin{equation}
    \hat{P}_m = \frac{1}{|B_m|}\sum_{i \in B_m}\hat{p}_i
\end{equation}}
{where $y_i$, $\hat{p}_i$, as defined in BS. Combining both equations, we get ECE as
\begin{equation}
    ECE = \frac{1}{N}\sum_m |B_m||\hat{\pi}_m-\hat{P}_m|
\end{equation}}
We report these diagnostics at both the waveform level and after cougher-level aggregation (mean probability per cougher), summarizing results as mean and standard deviation across outer folds, for models trained on audio-only and fused features.

\subsection{Evaluation Metrics}
In order to examine model performance in depth, we summarize performance using both threshold-dependent and threshold-free metrics. Let $y_i\in\{0,1\}$ denote the ground-truth label and $\hat{y}_i(\tau)=\mathbb{I}[\hat{p}_i\ge \tau]$ the predicted label obtained by thresholding the predicted probability $\hat{p}_i$ at $\tau\in[0,1]$, where 
\begin{equation}
\mathbb{I}[A] \;=\;
\begin{cases}
1, & \text{if } A \text{ is true},\\
0, & \text{if } A \text{ is false}.
\end{cases}
\end{equation}
is the indicator function. We define
\begin{align}
\mathrm{TP}(\tau)&=\sum_{i=1}^{N}\mathbb{I}[y_i=1,\hat{y}_i(\tau)=1] \\
\mathrm{FP}(\tau)&=\sum_{i=1}^{N}\mathbb{I}[y_i=0,\hat{y}_i(\tau)=1]
\end{align}
as True Positives and False Positives, respectively. Similarly, we define
\begin{align}
\mathrm{TN}(\tau)&=\sum_{i=1}^{N}\mathbb{I}[y_i=0,\hat{y}_i(\tau)=0] \\
\mathrm{FN}(\tau)&=\sum_{i=1}^{N}\mathbb{I}[y_i=1,\hat{y}_i(\tau)=0]
\end{align}
as True Negatives and False Negatives, respectively. From these we compute Sensitivity (Recall/True Positive Rate - TPR), Specificity (True Negative Rate - TNR), Positive Predictive Value (PPV), Negative Predictive Value (NPV), and Unweighted Average Recall (UAR) as
\begin{align}
\mathrm{Sens}(\tau)&=\frac{\mathrm{TP}(\tau)}{\mathrm{TP}(\tau)+\mathrm{FN}(\tau)} \\
\mathrm{Spec}(\tau)&=\frac{\mathrm{TN}(\tau)}{\mathrm{TN}(\tau)+\mathrm{FP}(\tau)} \\
\mathrm{PPV}(\tau)&=\frac{\mathrm{TP}(\tau)}{\mathrm{TP}(\tau)+\mathrm{FP}(\tau)} \\
\mathrm{NPV}(\tau)&=\frac{\mathrm{TN}(\tau)}{\mathrm{TN}(\tau)+\mathrm{FN}(\tau)} \\
\mathrm{UAR}(\tau)&=\frac{\mathrm{Sens}(\tau)+\mathrm{Spec}(\tau)}{2}
\end{align}
We also report Youden's index, which is useful for operating-point selection:
\begin{equation}
J(\tau)=\mathrm{Sens}(\tau)+\mathrm{Spec}(\tau)-1.
\end{equation}
To provide threshold-independent summaries, we report the area under the ROC curve (ROC AUC) and the area under the precision-recall curve (PR AUC). Let 
\begin{equation}
\mathrm{TPR}(\tau)=\mathrm{Sens}(\tau)
\end{equation}
and 
\begin{equation}
\mathrm{FPR}(\tau)=\frac{\mathrm{FP}(\tau)}{\mathrm{FP}(\tau)+\mathrm{TN}(\tau)}=1-\mathrm{Spec}(\tau)    
\end{equation}
The ROC curve is the parametric curve $\{(\mathrm{FPR}(\tau),\mathrm{TPR}(\tau)):\tau\in[0,1]\}$ and the ROC AUC is
\begin{equation}
\mathrm{AUC}_{\mathrm{ROC}}=\int_{0}^{1}\mathrm{TPR}(u)\,du,
\end{equation}
where $u=\mathrm{FPR}$. Similarly, the precision-recall (PR) curve is the parametric curve $\{(\mathrm{Recall}(\tau),\mathrm{Precision}(\tau)):\tau\in[0,1]\}$ with $\mathrm{Recall}(\tau)=\mathrm{Sens}(\tau)$ and $\mathrm{Precision}(\tau)=\mathrm{PPV}(\tau)$, and the PR AUC is
\begin{equation}
\mathrm{AUC}_{\mathrm{PR}}=\int_{0}^{1}\mathrm{Precision}(r)\,dr,
\end{equation}
where $r=\mathrm{Recall}$. In practice, these integrals are computed numerically from the finite set of thresholds induced by the model scores.

\subsection{Uncertainty Quantification}
Conformal prediction is used to quantify uncertainty of each model. We define the nonconformity score for binary probabilities as
\begin{equation}
s_i = 1 - p_{y_i}(x_i)    
\end{equation}
where 
\begin{align}
    p_{y_i = 1}(x) &=\hat p_{\mathrm{pos}}(x) \\
    p_{y_i = 0}(x) &=1-\hat p_{\mathrm{pos}}(x)
\end{align} 
For each target miscoverage level $\alpha$, we compute $\hat q_{\alpha}$ from the calibration scores via the order statistic
\begin{equation}
    k=\left\lceil (n+1)(1-\alpha)\right\rceil
\end{equation} 
and set $\hat q_{\alpha}=s_{(k)}$. Given a test point $x$, we output the conformal prediction set 
\begin{equation}
    \Gamma_{\alpha}(x)=\{1:\hat p_{\mathrm{pos}}(x)\ge 1-\hat q_{\alpha}\}\cup\{0:1-\hat p_{\mathrm{pos}}(x)\ge 1-\hat q_{\alpha}\}
\end{equation}
Because the TB label is defined \textit{per cougher}, conformal prediction should be evaluated at the same unit to satisfy its key assumption of \textit{exchangeability} between calibration points and future test points. Standard split conformal validity relies on the calibration examples and the test example being interchangeable draws from the same distribution. At the waveform level, however, multiple cough recordings from the same cougher are \textit{not} independent and typically share cougher-specific and session-specific factors (physiology, microphone/channel, environment, disease state, recording protocol), creating clustered dependence. Treating these correlated waveforms as separate calibration/test points violates exchangeability and can lead to miscalibrated coverage—often overly optimistic—because the effective sample size is closer to the number of coughers than the number of waveforms. In contrast, aggregating waveform evidence into a single cougher-level representation (e.g., mean probability) and applying conformal prediction on cougher-level examples aligns the statistical unit with the clinical decision (patient screening) and restores a much more plausible exchangeability condition across coughers, making the reported coverage, average set size, and singleton rate interpretable as distribution-free guarantees at the cougher level.
Thus, we compute $\hat q_{\alpha}$ and evaluate $\Gamma_{\alpha}$ only after cougher-mean aggregation  (using calibrated probabilities), reporting coverage, mean set size, and singleton rate on each outer test fold. It is reminded that coverage answers whether the output prediction sets can be trusted to contain the true label at the chosen rate ($1-\alpha$), mean set size answers whether these prediction sets are broad or narrow, on average, and how useful they can be, and singleton rate informs about how often a clear single-label prediction is obtained. 

In addition to standard conformal metrics (coverage and average set size), we evaluate the practical utility of conformal prediction as a \emph{selective} (reject-option) classifier at the couhgher level. For each outer fold, the base model produces calibrated TB probabilities for each held-out cougher and a point prediction is obtained using a fixed operating threshold selected by Youden’s (J) on the training side of the fold. In parallel, conformal prediction is applied on a disjoint cougher-level calibration subset to construct a prediction set ($C(x)\subseteq{\mathrm{TB}+,\mathrm{TB}-}$) at miscoverage level ($\alpha$). We then treat singleton sets ($|C(x)|=1$) as \emph{accepted} decisions and ambiguous (${ \mathrm{TB}+,\mathrm{TB}-}$) as \emph{deferred} cases, and compute (i) overall point accuracy, (ii) conditional accuracy restricted to singleton outputs, ($\mathrm{Acc}\mid\text{singleton}$), (iii) conditional accuracy on deferred cases, ($\mathrm{Acc}\mid\text{ambiguous}$), and (iv) the fraction of correct point predictions that were returned as singletons, $P(\text{singleton}\mid\text{correct})$. All quantities are computed per outer fold and summarized as mean $\pm$ standard deviation across folds (macro), with an additional pooled estimate obtained by aggregating all held-out coughers across folds.

\section{Results}
\label{experiments}
In this section, we report performance metrics on test sets for audio features and their fusion with clinical metadata. Cougher-level and waveform-level analyses are provided. Since the operating threshold $\tau$ is selected from calibrated probabilities on a disjoint calibration split, we next report calibration diagnostics (Brier score and ECE) to verify probability reliability before concluding this section by presenting conformal prediction results. 

\subsection{Classification Performance}
In what follows, we summarize classification performance obtained using audio-only features before moving to fused-feature results.

\subsubsection{Audio-only Features}
Table~\ref{tab:lrcb_acoustic} presents the classification results for both classifiers, LR and CatBoost, trained on acoustic features only. It is clear that both models achieve only moderate discrimination using this feature set, and most of the practical differences show up in the \textit{sensitivity–specificity trade-off} rather than in AUCs.
\begin{table*}[htb!]
\centering
\caption{\it Acoustic features only performance summary (mean $\pm$ std over outer folds) at waveform and cougher levels. Best per level in bold.}\label{tab:lrcb_acoustic}
\small
\begin{tabular}{lcccc}
\toprule
& \multicolumn{2}{c}{\textbf{Waveform} ($\mu\pm\sigma$)} & \multicolumn{2}{c}{\textbf{Cougher} ($\mu\pm\sigma$)} \\
\cmidrule(lr){2-3}\cmidrule(lr){4-5}
\textbf{Metric} & \textbf{LR} & \textbf{CatBoost} & \textbf{LR} & \textbf{CatBoost} \\
\midrule \midrule
Threshold ($\tau$) & $0.32\pm 0.04$ & $0.32\pm 0.05$ & $0.30\pm 0.04$ & $0.30\pm 0.05$ \\
ROC AUC & $0.68 \pm 0.05$ & $\mathbf{0.70 \pm 0.04}$ & $0.70 \pm 0.07$ & $0.70 \pm 0.07$ \\
PR AUC  & $0.47 \pm 0.07$ & $0.47 \pm 0.06$ & $0.45 \pm 0.10$ & $0.46 \pm 0.10$ \\
UAR     & $0.63 \pm 0.04$ & $\mathbf{0.65\pm 0.03}$ & $0.62 \pm 0.06$ & $\mathbf{0.64 \pm 0.05}$ \\
Sensitivity & $0.60 \pm 0.10$ & $\mathbf{0.68 \pm 0.11}$ & $0.63 \pm 0.19$ & $\mathbf{0.69 \pm 0.12}$ \\
Specificity & $\mathbf{0.67\pm0.07}$ & $0.61 \pm 0.11$ & $\mathbf{0.62\pm 0.13}$ & $0.58 \pm 0.12$ \\
PPV     & $\mathbf{0.44\pm0.04}$ & $0.43 \pm 0.04$ & $\mathbf{0.39 \pm 0.04}$ & $0.38 \pm 0.05$ \\
NPV     & $0.80 \pm 0.05$ & $\mathbf{0.82\pm 0.04}$ & $0.83 \pm 0.07$ & $\mathbf{0.84 \pm 0.05}$ \\
\bottomrule
\end{tabular}
\end{table*}
AUCs and UAR are meaningfully above chance, but they are more consistent with a \textit{weak-to-moderate screening signal} in cough acoustics alone. The modest PR AUC and PPV (roughly $0.38-0.44$) also suggest that, at the chosen operating point, a non-trivial fraction of TB+ predictions are false alarms. Moving from per-cough to per-cougher metrics does \textit{not} change ranking performance (ROC AUC stays similar), but it \textit{changes the operating behavior}: both models generally show slightly higher sensitivity and NPV after aggregation, paired with lower specificity and PPV. Intuitively, averaging over a cougher can stabilize noisy per-cough scores, but with the selected threshold it can also make the system more willing to flag TB (fewer false negatives and thus higher sensitivity/NPV) at the cost of more false positives (lower specificity/PPV). Nevertheless, it is expected when patient-level decisions are involved. CatBoost is \textit{consistently a bit more sensitivity-oriented} than LR in this setting. At waveform level, CatBoost improves ROC AUC ($0.70$ vs $0.68$) and UAR ($0.65$ vs $0.63$) mainly by boosting sensitivity ($0.68$ vs $0.60$), but with reduced specificity ($0.61$ vs $0.67$). The same pattern persists at cougher level: CatBoost keeps sensitivity around $0.69$, whereas LR is around $0.63$, but LR retains higher specificity. Hence, one can derive that CatBoost allows for more sensitivity but LR is more specific, and neither dominates across all metrics. This is not necessarily a modeling issue but most probably a feature issue. Cougher-level metrics often show \textit{larger standard deviations}, especially for sensitivity/specificity, because fewer independent units (coughers rather than cough segments) are evaluated per fold. 

\subsubsection{Fused Features}
Table~\ref{tab:lrcb_fused} presents the classification results for both classifiers, LR and CatBoost, trained on acoustic and clinical features combined.
\begin{table*}[htb!]
\centering
\caption{\it Fused features performance summary (mean $\pm$ std over outer folds) at waveform and cougher levels. Best per level in bold.}\label{tab:lrcb_fused}
\small
\begin{tabular}{lcccc}
\toprule
& \multicolumn{2}{c}{\textbf{Waveform} ($\mu \pm \sigma$)} & \multicolumn{2}{c}{\textbf{Cougher} ($\mu \pm \sigma$)} \\
\cmidrule(lr){2-3}\cmidrule(lr){4-5}
\textbf{Metric} & \textbf{LR} & \textbf{CatBoost} & \textbf{LR} & \textbf{CatBoost} \\
\midrule  \midrule
Threshold ($\tau$) & $0.31 \pm 0.05$ & $0.36 \pm 0.07$ & $0.33 \pm 0.08$ & $0.29 \pm 0.09$ \\
ROC AUC            & $0.80 \pm 0.04$ & $\mathbf{0.81 \pm 0.05}$ & $0.80 \pm 0.05$ & $\mathbf{0.81 \pm 0.05}$ \\
PR AUC             & $\mathbf{0.64 \pm 0.07}$ & $\mathbf{0.64 \pm 0.07}$ & $0.62 \pm 0.09$ & $\mathbf{0.64 \pm 0.08}$ \\
UAR                & $\mathbf{0.72 \pm 0.04}$ & $\mathbf{0.72 \pm 0.06}$ & $0.71 \pm 0.05$ & $\mathbf{0.73 \pm 0.05}$ \\
Sensitivity        & $\mathbf{0.73 \pm 0.11}$ & $0.72 \pm 0.16$ & $0.70 \pm 0.12$ & $\mathbf{0.78 \pm 0.12}$ \\
Specificity        & $\mathbf{0.71 \pm 0.09}$ & $\mathbf{0.71 \pm 0.12}$ & $\mathbf{0.73 \pm 0.12}$ & $0.67 \pm 0.14$ \\
PPV                & $\mathbf{0.53 \pm 0.07}$ & $0.52 \pm 0.06$ & $\mathbf{0.50 \pm 0.08}$ & $0.49 \pm 0.09$ \\
NPV                & $\mathbf{0.87 \pm 0.05}$ & $\mathbf{0.87 \pm 0.06}$ & $0.87 \pm 0.04$ & $\mathbf{0.90 \pm 0.04}$ \\
\bottomrule
\end{tabular}
\end{table*}

These results are obviously stronger than the acoustic-only ones. Both models achieve a ROC AUC of $0.80-0.81$ and a PR AUC of $0.62-0.64$, with UAR in the range of $0.71-0.73$. This is a clear indication that the \textbf{\textit{clinical variables contribute substantial complementary signal}}, helping disambiguate cough patterns that are otherwise non-specific. The PPV remains moderate ($0.49-0.53$), while NPV is consistently high ($0.87-0.90$), which fits the expected profile of a screening/triage tool: better at ruling out than definitively ruling in.
For LR, waveform- and cougher-level performance are very close in discrimination: ROC AUC is close to $0.80$ in both, with a small drop in PR AUC from $0.64$ to $0.62$ at cougher level. The main change is the operating trade-off: cougher aggregation slightly reduces sensitivity (from $0.73$ to $0.70$) but increases specificity (from $0.71$ to $0.73$). This suggests that averaging per cougher makes LR more conservative on detecting TB, yielding fewer false positives but more false negatives at the chosen threshold. NPV remains high ($0.87$), suggesting the model is relatively stable under aggregation, but the decision boundary shifts toward fewer referrals. For CatBoost, discrimination is again stable across levels (ROC AUC is around $0.81$ and PR AUC around $0.64$ in both), but aggregation pushes the operating point in the \textit{opposite} direction compared to LR. At cougher level, CatBoost becomes more sensitivity-oriented: sensitivity increases notably (from $0.72$ to $0.78$) while specificity decreases ($0.71$ to $0.67$). This is reflected in UAR improving slightly (from $0.72$ to $0.73$). Practically, cougher aggregation allows CatBoost to get fewer misses at the cost of more false alarms. This is often an acceptable trade in screening, depending on resource constraints for confirmatory testing.
The two classifiers are very similar in ranking performance, with CatBoost only marginally higher ROC AUC ($0.81$ vs $0.80$) and comparable PR AUC. The more meaningful difference is again at the operating point: LR (cougher level) leans toward higher specificity, while CatBoost (cougher level) leans toward higher sensitivity and NPV. It is evident that model choice should be guided less by AUC and more by the intended triage operating objective between minimization of missed TB (CatBoost) or unnecessary referral reduction (LR). It should be noted that an immediate conclusion across both models is that \textit{\textbf{multimodal} (acoustic and clinical)} modeling substantially improves TB screening performance relative to acoustic-only, and cougher-level aggregation mainly changes the sensitivity–specificity balance, which should be tuned to the screening context.

\subsection{Calibration Diagnostics}
Given that thresholding and uncertainty quantification rely on probabilistic outputs, we next evaluate probability calibration before presenting conformal prediction results. It is reminded that since waveform-level cough samples from the same cougher are not exchangeable, we apply conformal prediction only after aggregating to cougher-level probabilities, where exchangeability is more plausible and coverage/efficiency metrics are interpretable as cougher-level guarantees.

\subsubsection{Audio-only Features}
Calibration diagnostics for audio-only trained models are shown in Table~\ref{tab:calibration_diagnostics_lr_cb_audio}.
\begin{table*}[htb!]
\centering
\caption{\it Probability calibration diagnostics (mean $\pm$ std over outer folds) for Logistic Regression (LR) and CatBoost (CB) trained on audio-only features, before (raw) and after isotonic calibration, at waveform and cougher levels. }
\label{tab:calibration_diagnostics_lr_cb_audio}
\small
\begin{tabular}{l c c c c}
\toprule
& \multicolumn{2}{c}{\textbf{LR}} & \multicolumn{2}{c}{\textbf{CB}} \\
\cmidrule(lr){2-3}\cmidrule(lr){4-5}
\textbf{Metric} & \textbf{Raw} ($\mu \pm \sigma$) & \textbf{Isotonic} ($\mu \pm \sigma$) & \textbf{Raw} ($\mu \pm \sigma$) & \textbf{Isotonic} ($\mu \pm \sigma$) \\
\midrule \midrule
Waveform Brier score      & $0.22 \pm 0.02$ & $0.19 \pm 0.02$ & $0.20 \pm 0.01$ & $0.19 \pm 0.02$ \\
Waveform ECE              & $0.16 \pm 0.03$ & $0.07 \pm 0.01$ & $0.11 \pm 0.03$ & $0.06 \pm 0.02$ \\
\hline 
Cougher Brier score       & $0.22 \pm 0.02$ & $0.18 \pm 0.01$ & $0.20 \pm 0.02$ & $0.18 \pm 0.01$ \\
Cougher ECE               & $0.21 \pm 0.03$ & $0.11 \pm 0.03$ & $0.15 \pm 0.02$ & $0.11 \pm 0.03$ \\
\hline
\end{tabular}
\end{table*}
In this experiment, CatBoost produced more reliable probability estimates than LR prior to calibration, with lower waveform-level ECE ($0.11 \pm 0.03$ vs. 
$0.16 \pm 0.03$) and cougher-level ECE ($0.15 \pm 0.02$ vs. $0.21 \pm 0.03$), as well as lower Brier scores. Isotonic regression improved calibration for both models, reducing waveform-level ECE to $0.06 \pm 0.02$ (CatBoost) and $0.07 \pm 0.01$ (LR), and yielding nearly identical cougher-level ECE after calibration ($\approx 0.11$ for both). The larger improvement for LR reflects greater initial miscalibration, while the convergence of cougher-level calibration indicates that post-hoc calibration is crucial for producing interpretable cougher-level risk scores from cough acoustics alone, supporting thresholding and uncertainty-aware evaluation.

\subsubsection{Fused Features}
Calibration diagnostics for models trained on fused features are shown in Table~\ref{tab:calibration_diagnostics_lr_cb_fused}.
\begin{table*}[htb!]
\centering
\caption{\it Probability calibration diagnostics (mean $\pm$ std over outer folds) for Logistic Regression (LR) and CatBoost (CB) trained on fused features, before (raw) and after isotonic calibration, at waveform and cougher levels.}\label{tab:calibration_diagnostics_lr_cb_fused}
\label{tab:calibration_diagnostics_lr_cb_audio_fused}
\small
\begin{tabular}{l c c c c}
\toprule
& \multicolumn{2}{c}{\textbf{LR}} & \multicolumn{2}{c}{\textbf{CB}} \\
\cmidrule(lr){2-3}\cmidrule(lr){4-5}
\textbf{Metric} & \textbf{Raw} ($\mu \pm \sigma$) & \textbf{Isotonic} ($\mu \pm \sigma$) & \textbf{Raw} ($\mu \pm \sigma$) & \textbf{Isotonic} ($\mu \pm \sigma$) \\
\midrule
\midrule
Waveform Brier score   & $0.17 \pm 0.01$ & $0.16 \pm 0.02$ & $0.16 \pm 0.02$ & $0.16 \pm 0.02$ \\
Waveform ECE           & $0.11 \pm 0.04$ & $0.07 \pm 0.02$ & $0.10 \pm 0.02$ & $0.08 \pm 0.03$ \\
\hline
Cougher Brier score    & $0.17 \pm 0.02$ & $0.15 \pm 0.02$ & $0.16 \pm 0.02$ & $0.15 \pm 0.02$ \\
Cougher ECE            & $0.14 \pm 0.03$ & $0.10 \pm 0.02$ & $0.11 \pm 0.03$ & $0.10 \pm 0.02$ \\
\hline
\end{tabular}
\end{table*}
In this setting, both models produced substantially better-calibrated probabilities than in the audio-only setting, with raw Brier scores around $0.16-0.17$ and raw ECE values around $0.10-0.14$ at waveform and cougher levels. Isotonic regression consistently improved calibration, reducing waveform ECE from $0.11 \pm 0.04$ to $0.07 \pm 0.02$ for Logistic Regression and from $0.10 \pm 0.02$ to $0.08 \pm 0.03$ for CatBoost. At the cougher level, ECE decreased from $0.14 \pm 0.03$ to $0.10 \pm 0.02$ (LR) and remained stable to slightly improved for CatBoost ($0.11 \pm 0.02$ to $0.10 \pm 0.02$), while Brier scores decreased modestly for both models. Overall, \textit{fusing clinical inputs with cough acoustics yields more reliable risk estimates, and post-hoc calibration further improves probabilistic interpretability}, supporting consistent thresholding and uncertainty-aware evaluation at the cougher level.

\subsection{Uncertainty Quantification}
In the following paragraph, we quantify predictive uncertainty using conformal prediction, starting with the audio-only setting and concluding with the fused feature setting.

\subsubsection{Audio-only Features}
In Table~\ref{tab:lrcb_conformal}, we present conformal prediction outputs for models trained on acoustic features only.
\begin{table*}[htb!]
\centering
\caption{\it Conformal prediction results (mean $\pm$ std over outer folds) from models trained on acoustic features, at cougher level, for Logistic Regression (LR) and CatBoost (CB).}\label{tab:lrcb_conformal}
\small
\begin{tabular}{lccccc}
\toprule
& & \multicolumn{2}{c}{\textbf{LR}} & \multicolumn{2}{c}{\textbf{CB}} \\
\cmidrule(lr){3-4}\cmidrule(lr){5-6}
\textbf{Level} & $\alpha$ & \textbf{Coverage} ($\mu \pm \sigma$) & \textbf{Set size} ($\mu \pm \sigma$) \textbf{[Singleton]} & \textbf{Coverage} ($\mu \pm \sigma$) & \textbf{Set size} ($\mu \pm \sigma$) \textbf{[Singleton]} \\
\midrule \midrule
Cougher & $0.10$ & $0.90 \pm 0.04$ & $1.44 \pm 0.08\,[0.56]$ & $0.90 \pm 0.04$ & $1.43 \pm 0.10\,[0.57]$ \\
Cougher & $0.05$ & $0.95 \pm 0.03$ & $1.64 \pm 0.08\,[0.36]$ & $0.95 \pm 0.04$ & $1.60 \pm 0.09\,[0.40]$ \\
\bottomrule
\end{tabular}
\end{table*}
Cougher-level conformal prediction produced well-calibrated uncertainty sets for both LR and CatBoost, with empirical coverage closely matching the nominal targets across outer folds ($0.90$ for $\alpha=0.10$ and $0.95$ for $\alpha=0.05$, with modest fold-to-fold variability). As expected, tightening the error level from $\alpha=0.10$ to $\alpha=0.05$ increased the average prediction set size and reduced the singleton rate, reflecting the greater conservatism required to guarantee higher coverage. In this binary setting, the reported efficiencies imply that at $\alpha=0.10$ roughly $56-57\%$ of coughers receive a decisive singleton prediction, whereas at $\alpha=0.05$ only $36-40\%$ do so, with the remainder assigned the ambiguous set ({$\mathrm{TB+}, \mathrm{TB-}$}). Comparing models at similar coverage, CatBoost is marginally more efficient, yielding slightly smaller sets and higher singleton rates than LR, particularly under the stricter $\alpha=0.05$ value, suggesting it provides more informative conformal outputs without compromising validity.

\subsubsection{Fused Features}
Conformal prediction results for models trained on fused features are shown in Table~\ref{tab:cp_lrcb_fused}.
\begin{table*}[htb!]
\centering
\caption{\it Conformal prediction results (mean $\pm$ std over outer folds) from models trained on fused features, at cougher level, for Logistic Regression (LR) and CatBoost (CB).}\label{tab:cp_lrcb_fused}
\small
\begin{tabular}{lccccc}
\toprule
& & \multicolumn{2}{c}{\textbf{LR}} & \multicolumn{2}{c}{\textbf{CB}} \\
\cmidrule(lr){3-4}\cmidrule(lr){5-6}
\textbf{Level} & $\alpha$ & \textbf{Coverage} ($\mu \pm \sigma$) & \textbf{Set size} ($\mu \pm \sigma$) \textbf{[Singleton]} & \textbf{Coverage} ($\mu \pm \sigma$) & \textbf{Set size} ($\mu \pm \sigma$) \textbf{[Singleton]} \\
\midrule \midrule
Cougher & $0.10$ & $0.91 \pm 0.03$ & $1.32 \pm 0.08\,[0.68]$ & $0.91 \pm 0.04$ & $1.31 \pm 0.07\,[0.69]$ \\
Cougher & $0.05$ & $0.96 \pm 0.02$ & $1.51 \pm 0.07\,[0.49]$ & $0.96 \pm 0.02$ & $1.52 \pm 0.09\,[0.48]$ \\
\bottomrule
\end{tabular}
\end{table*}
Cougher-level conformal prediction remained well calibrated for both LR and CatBoost when moving from acoustic-only to fused (acoustic and clinical) features, with empirical coverage closely tracking the nominal targets and showing only small changes (acoustic-only features: ($\approx 0.90/0.95$) for ($\alpha=0.10/0.05$), while for fused features: ($\approx 0.91/0.96$)). The main effect of adding clinical features was a clear gain in \textit{efficiency}: prediction sets became substantially sharper (smaller average size) and more often decisive (higher singleton rate) at the same target coverage. For LR, the singleton rate increased from $0.56$ to $0.68$ at $\alpha=0.10$ and from $0.36$ to $0.49$ at $\alpha=0.05$, corresponding to average set sizes dropping from $1.44$ to $1.32$ and $1.64$ to $1.51$, respectively. CatBoost showed a similar pattern, with singleton rising from $0.57$ to $0.69$ at $\alpha=0.10$ and from $0.40$ to $0.48$ at $\alpha=0.05$, and set size decreasing from $1.43$ to $1.31$ and $1.60$ to $1.52$. In summary, fused features reduce the fraction of coughers assigned the ambiguous set ({$\mathrm{TB+},\mathrm{TB-}$}), particularly at $\alpha=0.10$, indicating that \textbf{\textit{clinical information makes the underlying risk estimates more separable and allows conformal prediction to preserve validity while issuing single-label decisions more frequently}}. Differences between LR and CatBoost under fused features were small relative to fold variability, suggesting that feature enrichment, rather than model choice, is the dominant driver of conformal efficiency in this setting.

\subsubsection{Selective Correctness}
Tables~\ref{tab:selcorr_audio_lr_cb} and~\ref{tab:selcorr_fused_lr_cb} present cougher-level selective performance under conformal prediction, comparing LR vs CatBoost trained on audio-only and fused features, at two miscoverage levels $\alpha=0.10,\: 0.05$. We treat singleton CP sets as accepted predictions and the ambiguous set $\{0,1\}$ as deferrals.
\begin{table*}[htb!]
\centering
\caption{\it Selective correctness metrics for audio-only models at the cougher level. Macro values are mean $\pm$ standard deviation across outer folds; pooled values aggregate all held-out coughers across folds.}
\small
\setlength{\tabcolsep}{4pt}
\begin{tabular}{ll|cc|cc|cc|cc}
\toprule
\multirow{2}{*}{Model} & \multirow{2}{*}{$\alpha$}
& \multicolumn{2}{c|}{Overall point accuracy}
& \multicolumn{2}{c|}{Acc$\mid$singleton}
& \multicolumn{2}{c|}{Acc$\mid$ambiguous}
& \multicolumn{2}{c}{P(singleton$\mid$correct)} \\
\cmidrule(lr){3-4}\cmidrule(lr){5-6}\cmidrule(lr){7-8}\cmidrule(lr){9-10}
& & Macro & Pooled & Macro & Pooled & Macro & Pooled & Macro & Pooled \\
\midrule \midrule
LR       & 0.10 & $0.62 \pm 0.06$ & 0.62 & $0.78 \pm 0.09$ & 0.78 & $0.42 \pm 0.05$ & 0.42 & $0.70 \pm 0.09$ & 0.70 \\
CatBoost & 0.10 & $0.61 \pm 0.07$ & 0.61 & $0.75 \pm 0.11$ & 0.74 & $0.43 \pm 0.06$ & 0.44 & $0.70 \pm 0.09$ & 0.70 \\
\midrule
LR       & 0.05 & $0.62 \pm 0.06$ & 0.62 & $0.84 \pm 0.08$ & 0.83 & $0.50 \pm 0.08$ & 0.51 & $0.48 \pm 0.10$ & 0.48 \\
CatBoost & 0.05 & $0.61 \pm 0.07$ & 0.61 & $0.86 \pm 0.12$ & 0.84 & $0.45 \pm 0.08$ & 0.47 & $0.55 \pm 0.10$ & 0.55 \\
\bottomrule
\end{tabular}
\label{tab:selcorr_audio_lr_cb}
\end{table*}
For \textit{audio-only} models, selective accuracy is strongly stratified by whether the conformal set is a singleton vs ambiguous, for both models: singleton outputs are substantially more reliable than ambiguous $\{0,1\}$ outputs. At $\alpha=0.10$, LR yields higher singleton accuracy (pooled $P(\mathrm{accuracy}\mid\text{singleton})=0.78$ vs.\ $0.74$ for CatBoost), whereas at $\alpha=0.05$ CatBoost is marginally higher ($0.84$ vs.\ $0.83$) and also converts a larger fraction of correct point decisions into singleton conformal outputs (pooled $P(\text{singleton}\mid\text{correct})=0.55$ vs.\ $0.48$ for LR). 
\begin{table*}[htb!]
\centering
\caption{\it Selective correctness metrics for fused (acoustic+clinical) models at the cougher level. Macro values are mean $\pm$ standard deviation across outer folds; pooled values aggregate all held-out coughers across folds.}
\label{tab:selcorr_fused_lr_cb}
\small
\setlength{\tabcolsep}{4pt}
\begin{tabular}{ll|cc|cc|cc|cc}
\toprule
\multirow{2}{*}{Model} & \multirow{2}{*}{$\alpha$}
& \multicolumn{2}{c|}{Overall point accuracy}
& \multicolumn{2}{c|}{Acc$\mid$singleton}
& \multicolumn{2}{c|}{Acc$\mid$ambiguous}
& \multicolumn{2}{c}{P(singleton$\mid$correct)} \\
\cmidrule(lr){3-4}\cmidrule(lr){5-6}\cmidrule(lr){7-8}\cmidrule(lr){9-10}
& & Macro & Pooled & Macro & Pooled & Macro & Pooled & Macro & Pooled \\
\midrule \midrule
LR       & 0.10 & $0.72 \pm 0.07$ & 0.72 & $0.84 \pm 0.07$ & 0.84 & $0.48 \pm 0.06$ & 0.47 & $0.79 \pm 0.06$ & 0.79 \\
CatBoost & 0.10 & $0.70 \pm 0.08$ & 0.70 & $0.80 \pm 0.11$ & 0.80 & $0.49 \pm 0.08$ & 0.49 & $0.79 \pm 0.04$ & 0.79 \\
\midrule
LR       & 0.05 & $0.72 \pm 0.07$ & 0.72 & $0.92 \pm 0.04$ & 0.92 & $0.54 \pm 0.09$ & 0.53 & $0.62 \pm 0.06$ & 0.62 \\
CatBoost & 0.05 & $0.70 \pm 0.08$ & 0.70 & $0.90 \pm 0.05$ & 0.90 & $0.52 \pm 0.12$ & 0.52 & $0.62 \pm 0.10$ & 0.61 \\
\bottomrule
\end{tabular}
\end{table*}
For \textit{fused (acoustic and clinical)} models, LR achieves higher singleton reliability (pooled $P(\mathrm{Acc}\mid\text{singleton})=0.84$ vs.\ $0.80$ at $\alpha=0.10$, and $0.92$ vs.\ $0.900$ at $\alpha=0.05$). CatBoost with fused features is slightly more decisive at $\alpha=0.10$ (singleton rate $0.69$ vs.\ $0.68$ for LR -- see Table~\ref{tab:cp_lrcb_fused}), but this comes with lower singleton accuracy, highlighting a modest trade-off between \emph{actionability} (singleton frequency, reported in Table~\ref{tab:cp_lrcb_fused}) and \emph{reliability} (singleton correctness) under conformal prediction.

\section{Discussion}
\label{discussion}
We structure the discussion around four aspects: the selected feature set, observed model performance, validation protocol, and uncertainty quantification.

\subsection{Feature Set}
The selected feature set is limited but well-known and suitable for an acoustic classification task, whereas it has been (partly or as a whole) used in literature for TB-detection from audio signals.
Our feature design emphasizes transparency and reproducibility while still capturing complementary acoustic cues from cough. We represent each cough using a compact set of short-time descriptors -- MFCCs (spectral envelope), chroma (pitch-class energy patterns), and simple spectral statistics (centroid, bandwidth, roll-off, flatness), computed on short overlapping frames and then summarized via distributional functionals (moments and percentiles) to obtain a fixed-length vector per recording. This approach intentionally balances interpretability, computational efficiency, and robustness: the features are widely available in standard toolkits, can be implemented consistently across studies, and yield stable tabular representations suitable for conventional ML models. It would be interesting, though, to research towards richer, higher-resolution, acoustic features coming from hand-crafted principles or complex model embeddings. 

\subsection{Model Performance}
Both models provide a strong, reproducible baseline with similar ranking performance (ROC AUC $\approx 0.80$; PR AUC $\approx 0.63$ with fused features). Model choice primarily affects the sensitivity--specificity trade-off at the operating point rather than overall discrimination. Specifically, CatBoost favors higher sensitivity/NPV at cougher level (fewer missed TB cases) while LR favors higher specificity (fewer false alarms). Across both LR and CatBoost, the primary performance driver is not the choice of classifier but the informational content of the input features. With acoustic-only features, both models achieve moderate discrimination and exhibit a clear operating-point trade-off, suggesting that the signal captured by hand-crafted cough descriptors is real but limited: differences between LR and CatBoost are mainly expressed in sensitivity–specificity balance rather than large gains in AUC. When clinical metadata are fused with acoustic features, performance increases substantially and consistently, indicating that routinely collected symptoms and demographics provide complementary information that helps resolve ambiguity in cough-only patterns. Importantly, the fused models maintain high NPV with only moderate PPV, aligning with the intended role of a screening/triage tool. These patterns reinforce that future performance improvements may come more from advanced representations (e.g., dynamics-aware audio modeling and richer fusion) and careful operating-point selection than from simply swapping model families. However, this study does not include late-decade developments in ML models, such as Convolutional Neural Networks and their variants, Recurrent Neural Networks (LSTM, GRU), and others, that can eliminate the need for hand-crafted features and are able to detect patterns in data that conventional ML models fail to do so.

\subsection{Validation}
A central contribution of this work is a validation protocol that prevents the most common source of optimistic bias in cough-based ML: leakage of subject (cougher) identity across splits. We adopt a cougher-disjoint nested cross-validation strategy, where an outer grouped split estimates generalization to unseen coughers, and an inner grouped split performs hyperparameter selection using only the proper-training coughers. Within each outer training pool, we further reserve a disjoint conformal-prediction-calibration subset at the cougher level to (i) select operating thresholds and (ii) compute conformal quantiles, while keeping the outer test fold fully untouched until final evaluation. This separation is critical because screening performance depends on operating-point choices: reporting sensitivity/specificity without a leakage-free thresholding procedure can materially overstate clinical utility. Finally, we report both waveform-level and cougher-level results, but interpret cougher-level outcomes as the clinically meaningful endpoint, since screening decisions are made per individual and waveform-level metrics can overweight subjects who contribute more coughs.

\subsection{Uncertainty}
Beyond point predictions, the pipeline explicitly quantifies confidence using a model-agnostic conformal prediction module, which converts calibrated probabilistic outputs into prediction sets with a user-selected coverage target under exchangeability. This is particularly relevant in cough-based screening, where uncertainty can be high due to heterogeneous symptoms and/or variable recording quality: a two-label prediction set can be interpreted as an ambiguous case that suggests retesting or confirmatory evaluation, whereas singleton sets support more decisive triage. Methodologically, we emphasize cougher-level conformal evaluation because the statistical guarantees are valid when the unit of analysis matches the unit of deployment (the patient) and when calibration/test examples are exchangeable. In other words, treating multiple coughs from the same cougher as independent points can violate these assumptions and yield misleading coverage. In our results, fused features not only improve discrimination but also improve conformal efficiency (smaller average set sizes and higher singleton rates at the same target coverage), indicating that \textit{adding clinical context makes model risk estimates more solid and uncertainty outputs more actionable for screening decisions}. 

It should be stressed that although the fused models reach high ROC-AUC values ($\approx 0.8$), the conformal prediction results indicate that this level of discrimination does not translate into consistently high decisiveness at the operating level. In particular, at a standard screening-style guarantee of $1-\alpha=0.90$, the average prediction-set size is still around $1.31-1.34$, implying that roughly one-third of cases receive a two-label set $(\{$TB+,TB-$\})$ (i.e., an output that can be labeled as uncertain) rather than a singleton decision. When the guarantee is tightened to $1-\alpha=0.95$, the average set size rises to $1.49-1.53$, and singleton rates drop to roughly one-half. Actually, this is expected: ROC-AUC summarizes \textit{ranking ability} across all thresholds, whereas conformal efficiency reflects how strongly separated the calibrated class-conditional score distributions are for individual samples. In a problem like TB cough screening, where symptoms are heterogeneous and recording conditions vary, a model can still rank positives above negatives reasonably well on average (yielding AUCs close to $0.8$) while producing many borderline scores near the decision boundary, which conformal prediction correctly flags as requiring abstention or follow-up. Practically, this highlights the value of uncertainty-aware screening: \textbf{\textit{performance should not be judged by AUC alone, and a substantial fraction of subjects may need confirmatory testing or repeat acquisition even when overall discrimination appears strong}}.

\subsection{Limitations}
Nevertheless, this study has several limitations. A first one concerns \textit{preprocessing and signal standardization}. In general, recordings are highly sensitive to recording hardware, distance-to-mic, room acoustics, and background noise. While distance-to-mic is fixed and signal-to-noise-ratio is relatively high in the CODA dataset, variability in all other aspects is acceptable for training a model that is sought to work in real-life settings. However, any fixed preprocessing pipeline (segmentation, denoising, amplitude normalization, silence trimming) risks either (i) removing diagnostically relevant information or (ii) leaving nuisance variability that the model can exploit spuriously. As a baseline study, our preprocessing choices prioritize reproducibility and thus, while some preprocessing schemes had been evaluated (trimming, noise removal, normalization) without improving outcomes, \textit{no preprocessing} has been applied to the cough waveform recordings.

A second limitation is \textit{model selection scope}. We focus on standard baselines (two models, a weak linear learner and a strong non-linear ensemble model) and a constrained hyperparameter search, which is appropriate for establishing a transparent reference but does not exhaust the modeling space. Alternative approaches, such as convolutional neural models that work on raw or close-to-raw input data, sequence models that capture cough temporal dynamics, self-supervised pretrained audio encoders, or multimodal architectures that learn joint representations of cough and clinical variables, could yield improvements, particularly with more data. Therefore, our results should be interpreted as strong reference points rather than state-of-the-art or upper bounds on achievable performance.

Third, \textit{feature set selection} may limit performance and robustness. Engineered time--frequency descriptors and distributional summaries compress cough signals into fixed-length vectors, which can be effective but may discard fine-grained temporal patterns (e.g., onset structure, cough phase transitions) that could be informative. Also, deep learning methods operating on spectrotemporal data have been shown---in a variety of audio datasets---to perform remarkably better than conventional ML models, limiting the need for domain knowledge for feature extraction. The fusion of audio and clinical features is also inherently heterogeneous: simple concatenation can underutilize cross-modal interactions compared to more sophisticated fusion methods. As a result, feature engineering remains a major determinant of outcomes, and the baseline feature set may not capture all clinically \mbox{relevant cues.}

Finally, \textit{uncertainty quantification assumptions} impose constraints. Conformal prediction provides coverage guarantees under exchangeability, but cough datasets often include correlated samples (multiple coughs per cougher, repeated sessions) and potential dataset shifts (different recording environments or populations). While we mitigate this by reporting cougher-level aggregation and by keeping calibration disjoint from test subjects, uncertainty outputs should still be treated as conditional on this specific dataset. 

\section{Conclusions and Future Work}
\label{conclusions}
In this work, we established a reproducible baseline for tuberculosis (TB) screening from cough audio augmented with clinical inputs, addressing a key obstacle in the field: the lack of standardized protocols that enable fair comparison across studies. We implemented a transparent end-to-end pipeline covering feature extraction, multimodal modeling with two representative baselines (Logistic Regression and CatBoost), and a consistent reporting suite of threshold-free (ROC-AUC, PR-AUC) and threshold-dependent (UAR, Sensitivity, Specificity, PPV, NPV) metrics. 

To ensure scientifically sound evaluation, we adopted a cougher-independent nested cross-validation strategy that separates hyperparameter tuning from final testing, and we further introduced a disjoint calibration subset within each outer fold to select operating thresholds and to construct conformal prediction sets. This design prevents information leakage, yielding performance estimates that better reflect generalization to unseen individuals and supporting both waveform-level and clinically relevant cougher-level reporting.

Beyond point estimates, we incorporated post-hoc probability calibration and a model-agnostic approach to quantify uncertainty, named conformal prediction, in a distribution-free manner. The resulting prediction sets provide an actionable mechanism for screening workflows: confident singleton outputs can support automated triage, while non-singleton outputs naturally indicate ambiguous cases suitable for re-recording or confirmatory testing. Overall, the presented protocol and results provide a common reference point that reduces methodological variance and enables future work to evaluate improvements attributable to modeling rather than to differences in data handling and evaluation.

Finally, we emphasize that the proposed pipeline suggests a \textit{screening} rather than a diagnostic tool: it aims to prioritize individuals for confirmatory evaluation. By releasing a standardized experimental protocol and strong baselines, this work supports more rigorous benchmarking and more cumulative progress toward deployable, uncertainty-aware TB screening from cough audio in real-world settings. 

Building on this standardized baseline, several directions can further improve robustness, comparability, performance, and clinical relevance. First, preprocessing can be strengthened by integrating and validating dedicated cough event detection and quality-control modules (e.g., silence removal) and by systematically studying sensitivity to segmentation. Second, {the proposed pipeline does not prescribe how features are constructed but instead provides a common benchmark that accommodates both conventional feature engineering and learned representations produced by modern deep learning. By holding the validation protocol and reporting standards fixed while allowing the modeling choices to vary, future work can innovate on architectures and representations while remaining directly comparable.} A transition to convolutional or recurrent neural models on spectrotemporal input representations seems a plausible direction, parallel to pretrained or self-supervised audio encoders and temporal models that preserve cough dynamics may capture information that is discarded by fixed summary statistics. Third, multimodal modeling can be expanded beyond early fusion by exploring late/attention-based fusion and modality-aware training that handles clinical variables while improving the interpretability of cross-modal interactions. From an evaluation standpoint, future work should emphasize clinically aligned operating points (e.g., sensitivity at a fixed specificity, or vice versa) alongside discrimination metrics. The conformal framework enables actionable triage policies; that is, a natural extension is to evaluate selective prediction (abstention/referral when prediction sets are non-singleton) and quantify the trade-off between referral rate and missed TB cases. Finally, external validation on independent cohorts and explicit domain-shift experiments (site/population) are essential to establish generalization beyond the current dataset. Coupled with open, fixed splits and a benchmark suite, these steps would enable more rigorous, cumulative progress toward deployable TB screening systems.

\section*{Acknowledgments}
The 
 datasets used for the analyses described were contributed by Adithya 
 Cattamanchi at UCSF and Simon Grandjean Lapierre at University of Montreal and were generated in collaboration with researchers at Stellenbosch University (Grant Theron), Walimu (William Worodria and Alfred Andama); De La Salle Medical and Health Sciences Institute (Charles Yu), Vietnam National Tuberculosis Program (Nguyen Viet Nhung), Christian Medical College (DJ Christopher), Centre Infectiologie Charles Mérieux Madagascar (Mihaja Raberahona and Rivonirina Rakotoarivelo), and Ifakara Health Institute (Issa Lyimo and Omar Lweno), with funding 
 from the U.S. National Institute of Health (U01 AI152087), The Patrick J. McGovern Foundation and Global Health Labs.

\section*{Material}
Source code and supplementary material can be found in
\begin{center}
    \url{https://github.com/gpk1983/TB-Screening-from-Cough-Audio}
\end{center}

\bibliographystyle{unsrtnat}
\bibliography{references}  

\end{document}